\documentclass[aps,prd,showpacs,superscriptaddress,nofootinbib,10pt]{revtex4-2}
\usepackage{acronym}
\usepackage{amsfonts}
\usepackage{amsmath}

\usepackage{amssymb}
\usepackage{bm}
\usepackage{cases}
\usepackage{color}
\usepackage{comment}
\usepackage[dvipsnames]{xcolor}
\usepackage{enumerate}
\usepackage{epsfig}
\usepackage{etoolbox}
\usepackage{fancyref}
\usepackage{fancyhdr}
\usepackage[T1]{fontenc} 
\usepackage{graphicx}
\usepackage{hyperref}
\usepackage{indentfirst}
\usepackage{latexsym}
\usepackage{mathrsfs}
\usepackage{mciteplus}
\usepackage{multirow}
\usepackage{rotating}
\usepackage{scrextend}
\usepackage{soul}
\usepackage{subfigure}
\usepackage{verbatim}
\usepackage{amsthm} 
\usepackage{tabularx}
\usepackage{enumitem}    
\usepackage{array}

\begin{document}

\title{Master functions and hybrid quantization of perturbed nonrotating black hole interiors}

\author{Michele Lenzi\,
}
\email{michele.lenzi@ulb.be}
\affiliation{Physique Théorique et Mathématique, Université Libre de Bruxelles (ULB), Campus Plaine, Building NO, CP 231, Boulevard du Triomphe B-1050, Bruxelles, Belgique}

\author{Guillermo A. Mena Marug\'an\,
}
\email{mena@iem.cfmac.csic.es}
\affiliation{Instituto de Estructura de la Materia, IEM-CSIC, C/ Serrano 121, 28006 Madrid, Spain}
\author{Andr\'es M\'{\i}nguez-S\'anchez\,
}
\email{andres.minguez@iem.cfmac.csic.es}
\altaffiliation{Affiliated to the PhD Program, Departamento de F\'isica Te\'orica, Universidad Complutense de Madrid, 28040 Madrid, Spain}
\affiliation{Instituto de Estructura de la Materia, IEM-CSIC, C/ Serrano 121, 28006 Madrid, Spain}

\begin{abstract}
Master functions of black holes are fundamental tools in gravitational-wave physics and are typically derived within a Lagrangian framework. Starting from the Kantowski-Sachs geometry, one can instead construct a perturbative Hamiltonian description for the interior region of an uncharged and nonrotating black hole. This approach provides a complementary perspective and enables a quantum treatment of the background geometry and its perturbations. In this work, we extend the application of this formulation to the exterior region and establish a correspondence between the perturbative invariants of the canonical approach and the master functions commonly used in black hole analyses. Once a consistent Hamiltonian description for their canonical counterparts is obtained, a hybrid quantization of the master functions follows naturally.
\end{abstract}

\maketitle

\section{Introduction \label{sec: I}}

Black holes (BHs) have become central objects in modern physics and astrophysics. Over the past decades, they have evolved into powerful laboratories for testing the fundamental laws of general relativity. Notable milestones include the detection of gravitational waves by LIGO and Virgo \cite{LIGO} and the first images of BH shadows captured by the Event Horizon Telescope \cite{EHT}. These achievements have transformed BHs into observable systems that may provide access to the strong-field regime of gravity. In parallel with these observational advances, theoretical studies about their perturbations and the relation with gravitational waves \cite{RW,Z,M,T} have gained renewed importance (see e.g. Ref. \cite{LISA2}). As the quantity and quality of measurements continue to improve, the need for refined models becomes increasingly evident \cite{BCS}. Among the many aspects that can be explored, perturbations help assess BH stability and, as we have indicated, play a crucial role in gravitational-wave physics.

Mergers of compact objects constitute one of the primary sources of currently detectable gravitational waves \cite{LIGO}. In the final stage of the merger, the newly formed BH approaches equilibrium and can be accurately described within perturbation theory \cite{LISA2,Nollert}. After a transient, the perturbed BH reaches a so-called ringdown phase, in which the gravitational radiation can be described in terms of quasinormal modes (QNMs) \cite{Nollert}. Moreover, these QNMs, which characterize the oscillations of the perturbed spacetime, depend on intrinsic parameters such as the mass, charge, and angular momentum of the BH \cite{BCS}. At present, detected events typically involve a range of masses that limit the accessible QNMs. However, in the near future, with the development of a new generation of gravitational-wave detectors, a much broader spectrum of ringdown signals is expected to become observable (see e.g. Ref. \cite{LISA}). Spectral measurements will enable stringent tests of general relativity, potentially revealing deviations that might point toward new physics.

Since the advent of quantum mechanics and general relativity, the quest for a unified description has been a central goal of physics. Numerous proposals have been put forward \cite{A&L,Thiemann,ST,AS}. One of them, which only aims at quantizing general relativity without unifying all interactions, is loop quantum gravity (LQG) \cite{A&L,Thiemann}. Many of our comments in this work will be especially suited for this quantum theory of gravity. However, despite the effort devoted to describe gravity as a quantum field, progress has been hindered by the lack of observational evidence of quantum gravity phenomena. Scenarios in which both the quantum and the gravitational frameworks are simultaneously relevant are rare and often difficult to probe. The earliest moments of the universe provide a clear example. The application of LQG to study the early universe has led to the development of a discipline called loop quantum cosmology (LQC) \cite{APS1,APS2,AP,ABRev}. In principle, another possible candidate is the merger of (supermassive) BHs, a highly energetic process where quantum effects might emerge during the transition from coalescence to equilibrium, leaving potential signatures detectable by next-generation interferometers. Although naive estimates could indicate that quantum gravity effects on the gravitational radiation emitted during such processes would not be significant, a careful analysis, including not only the propagation on an effective background corrected by quantum modifications but a full quantization of both the geometry and its perturbations investigating its physical consequences, has not been completed to date.

In fact, to address these questions, a new formalism for uncharged and nonrotating BHs was recently developed. The novelty of this approach lies in reformulating BH perturbations within a Hamiltonian framework, which extends earlier ideas \cite{ADM,DB1,DC} and incorporates more recent advances \cite{MM,MM2}. Among its appealing features, the framework offers a systematic and coordinate-independent method to identify the physical perturbations (perturbative gauge invariants) and supports diverse applications, such as symmetry analyses (notably in the context of Darboux transformations \cite{CM1,CM2,MGAC}) and the study of quantum aspects through the quantization of the full perturbed spacetime.

Based on a Kantowski-Sachs interior geometry \cite{K&S,E.Weber}, the BH background can be described by two canonical pairs subject to a Hamiltonian constraint. In terms of triad and connection variables, these pairs encode the (interior) spatial components of the geometry. For the quantum description, adopting LQC-inspired variables enables a well-defined representation of background quantities as essentially self-adjoint operators, leading to the resolution of the interior singularity and to an effective behavior that incorporates quantum deviations from Einstein theory \cite{A&B,AOS,AOS2,AO,BH_Con,BH_GAB,BH_Cong,GBA}. 

First-order perturbative dynamical degrees of freedom arise by expanding the perturbations of the metric in a basis that separates the angular, radial, and temporal dependencies (for simplicity, we use here a terminology for the radial and temporal directions based on the interior geometry, but this terminology is simply a convention). Modes are then classified into axial and polar (odd and even parity) contributions. Each mode gives rise to six perturbative canonical pairs (two axial and four polar), four linear perturbative constraints (one axial and three polar, that can be treated as gauge constraints, all of them corresponding to the linearization of the spacetime diffeomorphism constraints), and a single perturbative Hamiltonian (quadratic in perturbations, which can be considered as the contribution to the zero-mode of the Hamiltonian constraint on the combined system formed by the background and the perturbations) \cite{MM,MM2}. By performing suitable canonical transformations and redefining the background variables and Lagrange multipliers to include perturbative corrections \cite{Lan,LMM,MM,MM2}, this set can be reorganized into two gauge invariant pairs (one axial and one polar) and four additional independent pairs (one axial and three polar), each provided by a perturbative (gauge) constraint and its conjugate variable. This rearrangement not only identifies the physical perturbations but also keeps the gauge unfixed and a genuinely dynamical background \cite{DC,MM,MM2}. Moreover, from this structure a privileged Fock representation (up to unitary equivalence) can be defined for the gauge invariant perturbations (or, alternatively, for test fields on our dynamical background) \cite{AT}.

However, previous results were restricted to the BH interior \cite{MM,MM2} and, what is even worst, no direct connection to the master variables commonly used in the literature was made explicit \cite{BCS}. In this work, we propose a natural extension of the formalism by considering a complex canonical transformation of the background variables. As a result, a geometric description of the exterior region is obtained with the distinctive feature that the Hamiltonian interpretation transitions from an interior time evolution to an exterior radial evolution. The resulting framework provides a unified treatment of BHs (and their perturbations) in which the emphasis lies on the canonical variables rather than on the interior-exterior distinction. We show that this transformation extends straightforwardly to the perturbative level, where an additional canonical transformation (implemented through several intermediate steps for clarity) enables the perturbative gauge invariants to be rewritten as mode coefficients of generalized master functions. When the background is specialized to the Schwarzschild geometry, the Cunningham-Price-Moncrief (CPM) and Zerilli-Moncrief (ZM) invariants \cite{RW,Z,M} are recovered for the axial and polar sectors, respectively. In this respect, our work completes the results of Ref. \cite{MGAC}, where the connection to axial parity master functions was obtained, to the quite more complicated polar sector, finding the polar gauge invariants in terms of the ZM master function. The generality of our approach also opens the possibility of exploring other solutions and effective models arising from modifications of general relativity. Since all relevant expressions depend only on phase-space variables, such extensions simply require substituting the background functions (replacing in this way classical background solutions with their effective values), and analyzing then the resulting dynamics.

The quantization of the master functions builds on hybrid loop quantum cosmology (hybrid-LQC), which has proven to be a robust framework for systems with perturbations; see Refs. \cite{AAN1,AAN2,FMO,hyb1,hyb-review,hyb-others} for cosmological applications and its extensions to other geometries in Refs. \cite{IJV,KBP,Z&W}. Its versatility stems from a simple premise: background and perturbative degrees of freedom can be quantized using different methods. In regimes where quantum-gravity effects are significant, the dominant corrections are expected to originate from the background geometry, while perturbations contribute to them at a subdominant level. Even with this hierarchy, the combined quantization of the background and the perturbation is not trivial, because their consistency is put to the test by the zero-mode of the Hamiltonian constraint, which includes contributions from the two subsystems.

Motivated by these considerations, we adopt here a hybrid quantization scheme in which the background is quantized using an LQC representation, whereas the master functions are treated with standard quantum field techniques. The simplicity of the potentials (particularly the Regge-Wheeler potential) ensures that the quantum representation of all perturbative constraints is well defined under reasonable assumptions. Effective equations for the master functions, including quantum corrections, can then be derived \cite{hyb-review,BAl}. Remarkably, the hybrid quantization that we develop in this work can be straightforwardly generalized to other quantizations of the background different from that of LQC, as far as they satisfy some basic requirements, such as the resolution of the essential singularity at the center of the interior region of the BH and the invertibility of certain related geometric operators. 

The paper is organized as follows. Section \ref{sec: II} reviews the Kantowski-Sachs background and its perturbations, extending the formalism to the exterior region. Section \ref{sec: III} constructs the perturbative gauge invariants and establishes their relation to generalized master functions. The hybrid quantization of the system is discussed in Sec. \ref{sec: IV}. Finally, Sec. \ref{sec: V} summarizes the main conclusions and outlines future directions for research. Four appendices collect lengthy expressions relevant to the Hamiltonian formulation. Throughout this work, we adopt Planck units, with the speed of light and the reduced Planck and Newton constants equal to one.

\section{Hamiltonian formulation for linear perturbations of nonrotating black holes \label{sec: II}}

In this section, we review the Hamiltonian approach used for the BH background and its perturbations. The description is organized into three main parts: (i) the definition of the canonical pairs; (ii) the identification of the constraints; and (iii) the introduction (when necessary) of the Hamiltonian itself. More details on the foundational framework can be found in Refs. \cite{ADM,DC}, while comprehensive analyses of its application to BHs are provided in Refs. \cite{MM,MM2}. Here, we summarize only the main expressions relevant for our construction. 

\subsection{Background \label{ssec: II.A}}

Following previous works \cite{AOS2, AO, GBA}, we begin by presenting a Hamiltonian formulation based on the Kantowski-Sachs geometry for the BH interior. As a novel contribution, we then show how this framework naturally extends to the exterior region. For this analysis, the background canonical pairs are defined within a triad-connection formulation, mainly motivated by quantum considerations \cite{A&B,AOS}. Nevertheless, equivalent classical results can be obtained using other canonical variables. Specifically, the densitized triad $E$ and its Ashtekar-Barbero $SU(2)$ connection $A$ adopt the following expressions:
\begin{equation}
    \label{eq: II-2.1}
    E = p_c \text{sin}\theta \tau_3\partial_{\zeta} + \frac{p_b}{L_o} \text{sin}\theta \tau_2\partial_{\theta} - \frac{p_b}{L_o} \tau_1\partial_{\phi}, \qquad
    A = \frac{c}{L_o}\tau_3\text{d}\zeta + b\tau_2\text{d}\theta - b\text{sin}\theta \tau_1\text{d}\phi + \text{cos}\theta\tau_3\text{d}\phi.
\end{equation}
Here, $\zeta$ denotes the interior radial coordinate, namely the coordinate that parametrizes the set of orbits of spherical symmetry on the spatial sections of the interior region. Meanwhile, $\theta$ and $\phi$ are the usual angular coordinates on each of these orbits of spherical symmetry. The interior time coordinate will be denoted by $\chi$, this choice of notation will be explained in more detail below. The functions $p_b$ and $p_c$, that will be referred as triad variables, are time dependent and determine $E$. Conversely, the time dependent functions $b$ and $c$, referred to as connection variables, determine $A$. The $\tau_i$ (with $i=1,2,3$) are the standard basis elements in the algebra $\mathfrak{su}(2)$, satisfying the commutation relations $[\tau_i,\tau_j]=\epsilon_{ij}^{\;\;\;k}\tau_k$. Lastly, $L_o$ is a fiducial length introduced to compactify the $\zeta$--direction and therefore the spatial sections, avoiding in this way infrared subtleties related to infinite volumes \cite{AO}. 

Background symmetries considerably simplify the formulation. After performing the spatial integration, the action in Hamiltonian form and in triad-connection variables becomes
\begin{equation}
    \label{eq: II-2.2}
    S_0 = \int \bigg( \frac{1}{\gamma}p_b\text{d}b + \frac{1}{2\gamma}p_c\text{d}c - \tilde{H}_{\text{KS}}[\tilde{N}]\text{d}\chi \bigg),
\end{equation}
where $\gamma$ is the real Immirzi parameter, that reflects an ambiguity in the possible choices of connection variables \cite{A&L,Thiemann}. This model represents a fully constrained system with Poisson brackets given by $\{b,p_b\}_{\text{B}} = \gamma$ and $\{c,p_c\}_{\text{B}} = 2\gamma$. The dynamics arise from the following background Hamiltonian constraint:
\begin{equation}
    \label{eq: II-2.3}
    \tilde{H}_{\text{KS}}[\tilde{N}] = -\tilde{N}\frac{L_o}{2}\bigg( \Omega_b^2 + \frac{p_b^2}{L_o^2} + 2\Omega_b\Omega_c \bigg).
\end{equation}
Here, $\Omega_b = bp_b/(\gamma L_o)$ and $\Omega_c = cp_c/(\gamma L_o)$ are the generators of dilations in the corresponding $b$-- and $c$--sectors of phase space \cite{BH_GAB}. Hence, we will call them dilation functions in the following. On the other hand, note that one may redefine $b$ and $c$ to absorb $\gamma$ (case in which $b$ and $c$ can be identified as pure extrinsic-curvature variables), and set $L_o$ equal to one for further simplification. We will nonetheless keep both constants in our discussion for future convenience (for example, to allow the study of the uncompactifed or classical limits). Finally, the subscript $\text{B}$ indicates that the Poisson brackets refer to background variables, whereas $\tilde{N}$ is the densitized lapse function \cite{BH_Con,GBA}. This change of density weight merely corresponds to a different parametrization of the Hamiltonian evolution and does not alter the physics of the classical dynamics.

With these expressions, the interior geometry is fully determined. Our aim, however, is to adopt a broader viewpoint that places the interior and exterior descriptions on the same footing. For this purpose, we rewrite the background spacetime metric in terms of the triad variables as
\begin{equation}
    \label{eq: II-2.4}
    \text{d}s^2 = \frac{p_b^2}{L_o^2}\bigg(-|p_c| \tilde{N}^2 \text{d}\chi^2 + \frac{1}{|p_c|} \text{d}\zeta^2\bigg) + |p_c|(\text{d}\theta^2 + \sin^2\theta\text{d}\phi^2).
\end{equation}
Our notation for the coordinates $\chi$ and $\zeta$ wants to prevent us from unconsciously interpreting any of them directly as a \textit{time} or a \textit{radius}, which would be misleading for the type of analysis carried out here. Instead, $(\chi,\zeta)$ are treated as coordinates on a reduced two-dimensional submanifold of the BH spacetime, namely the manifold corresponding to the set of orbits of spherical symmetry. In this context, the complex\footnote{If a complex transformation appears cumbersome, an equivalent result follows in a metric formulation by flipping the sign of the metric function $p_b^2$ and of its conjugate momentum, avoiding imaginary terms.} canonical transformation,
\begin{equation}
    \label{eq: II-2.5}
    \bar{b} = -ib, \qquad \bar{p}_b = ip_b, \qquad \bar{c} = c, \qquad \bar{p}_c = p_c,
\end{equation}
slightly modifies the Hamiltonian constraint but has important conceptual consequences (see also Ref. \cite{AOS2} for a detailed discussion in an effective model). It exchanges the physical roles and spacelike/timelike character of $(\chi,\zeta)$ through a redefinition of the metric functions rather than through a coordinate transformation. In doing so, the resulting description corresponds to the exterior geometry. This reinterpretation makes clear that spherical symmetry and staticity (or, equivalently, surface orthogonality) reduce the BH spacetime to an effectively two-dimensional submanifold that naturally admits a canonical formulation. The phase space can be represented either by $(b,c,p_b,p_c)$ or by $(\bar{b},\bar{c},\bar{p}_b,\bar{p}_c)$, related through Eq. \eqref{eq: II-2.5}, which provide two equally valid formulations. In the first, the evolution parameter associated with the Hamiltonian constraint plays the role of a time variable; in the second, it acquires (up to an irrelevant sign) a radial interpretation, a direct consequence of how the canonical transformation acts on the spacetime metric. On the other hand, it is worth pointing out that the dynamics of our background depends on $b$ and $p_b$ only through the functions $\Omega_b$ and $p_b^2$. Both remain real under the considered complex transformation. Actually, $\Omega_b$ apparently remains invariant, while $p_b^2$ flips sign, from positive no negative. To this extent, we can avoid the use of the overbar in the transformed phase-space variables if we suitably allow for different ranges for $p_b^2$ before and after the transformation. We proceed in this way in the following, avoiding unnecessary complications in our notation.

A concrete example, later used to connect the perturbative results with QNMs, is the Schwarzschild background. It arises as a solution of the equations of motion for $\tilde{N} = -L_o^2/p_b^2$. After straightforward computations, we obtain
\begin{equation}
    \label{eq: II-2.6}
    \frac{p_b^2}{L_o^2} = -\chi^2f, \qquad |p_c| = \chi^2, \qquad \Omega_b = \chi f, \qquad \Omega_c = M,
\end{equation}
where $f(\chi) = 1 - 2M/\chi$ is the standard Schwarzschild function, with $M$ the ADM mass of the BH. The expressions hold for $\chi<2M$, corresponding to the interior description. The exterior solution follows directly from the complex canonical transformation and applies for $\chi>2M$. Together, both regions reproduce the familiar Schwarzschild metric,
\begin{equation}
    \label{eq: II-2.7}
    \text{d}s^2 = - f\text{d}\zeta^2 + \frac{1}{f}\text{d}\chi^2 + \chi^2(\text{d}\theta^2 + \sin^2\theta\text{d}\phi^2).
\end{equation}
Since the evolution parameter is always $\chi$, this example clearly illustrates that the interior formulation corresponds to a time evolution, whereas the exterior describes a radial evolution. Throughout this work, we retain a general background formulation, specializing to the Schwarzschild case only when an explicit comparison is required.

\subsection{Perturbations \label{ssec: II.B}}

The perturbations, though structurally more involved, can be treated in parallel with the background by defining the canonical pairs within a metric formulation. Since its choice is not as critical for the quantum study as in the background case \cite{hyb-review}, this selection can be made to facilitate the future comparison with the BH master functions \cite{BCS}. As is customary, the perturbations can be expanded in a mode series to simplify the analysis. More concretely, assuming that the background already contains any possible zero-mode contribution of the perturbations, the rest of linear contributions of the three-metric perturbation $h$, its conjugate momentum $p$, the lapse-function perturbation $C$, and the shift-vector perturbation $B$ (all of them with respect to the natural 3+1 decomposition of the spacetime metric in the interior region) provide the necessary ingredients for the Hamiltonian description and can be written as
\begin{equation}
    \label{eq: II-2.8}
    h = \sum_{\mathfrak{n},\lambda}h^{\mathfrak{n},\lambda}, \qquad p = \text{det}(E)\sum_{\mathfrak{n},\lambda}\sqrt{\frac{L_o^2}{p_b^2|p_c|}}p^{\mathfrak{n},\lambda}, \qquad C = \kappa\sum_{\mathfrak{n},\lambda}\sqrt{\frac{p_b^2|p_c|}{L_o^2}}C^{\mathfrak{n},\lambda} \qquad B = \kappa\sum_{\mathfrak{n},\lambda}B^{\mathfrak{n},\lambda},
\end{equation}
where $\kappa=16\pi$ in our chosen units. These background-dependent prefactors are introduced to streamline calculations involving density weights \cite{MM,MM2}. In Eq. \eqref{eq: II-2.8} each sum is a mode expansion corresponding to a real Fourier basis in the coordinate $\zeta$ and a real Regge-Wheeler-Zerilli spherical harmonic basis for the angular dependence on the sphere \cite{MM}, with expansion coefficients that depend dynamically on the coordinate $\chi$. More in detail, we have
\begin{equation}
    \label{eq: II-2.9}
    \begin{aligned}
        &\begin{aligned}
            h^{\mathfrak{n},\lambda} &= h_6^{\mathfrak{n},\lambda}Y^{lm}Q_{n,\lambda}\text{d}\zeta^2 - 2(h_1^{\mathfrak{n},\lambda}X^{lm}_A-h_5^{\mathfrak{n},\lambda}Z^{lm}_A)Q_{n,\lambda}\text{d}\zeta\text{d}x^A\\
            &+ (h_2^{\mathfrak{n},\lambda}X^{lm}_{AB} + h_3^{\mathfrak{n},\lambda}Y^{lm}_{AB} + h_4^{\mathfrak{n},\lambda}Z^{lm}_{AB})Q_{n,\lambda}\text{d}x^A\text{d}x^B,
        \end{aligned}\\
        &\begin{aligned}
            p^{\mathfrak{n},\lambda} &= \frac{p_b^4}{L_o^4p_c^2}p_6^{\mathfrak{n},\lambda}Y^{lm}Q_{n,\lambda}\text{d}\zeta^2 - \frac{1}{l(l+1)}\frac{p_b^2}{L_o^2}(p_1^{\mathfrak{n},\lambda}X^{lm}_A-p_5^{\mathfrak{n},\lambda}Z^{lm}_A)Q_{n,\lambda}\text{d}\zeta\text{d}x^A\\
            &+ 2p_c^2\frac{(l-2)!}{(l+2)!}\bigg(p_2^{\mathfrak{n},\lambda}X^{lm}_{AB} + \frac{1}{4}\frac{(l+2)!}{(l-2)!}p_3^{\mathfrak{n},\lambda}Y^{lm}_{AB} + p_4^{\mathfrak{n},\lambda}Z^{lm}_{AB}\bigg)Q_{n,\lambda}\text{d}x^A\text{d}x^B,
        \end{aligned}\\
        &C^{\mathfrak{n},\lambda} = - \frac{\tilde{N}}{2} k_3^{\mathfrak{n},\lambda}Y^{lm}Q_{n,\lambda},\\
        &B^{\mathfrak{n},\lambda} = \frac{p_b^2}{L_o^2} k_2^{\mathfrak{n},\lambda}Y^{lm}Q_{n,\lambda}\text{d}\zeta - (k_0^{\mathfrak{n},\lambda}X^{lm}_A -|p_c|k_1^{\mathfrak{n},\lambda}Z^{lm}_A)Q_{n,\lambda}\text{d}x^A.
    \end{aligned}
\end{equation}
The functions $Q_{n,\lambda}$ are sines and cosines which form a real Fourier basis, with $n\in\mathbb{N}_0$ and $\lambda\in\{+,-\}\simeq\mathbb{Z}_2$ where $\mathbb{N}_0$ denotes the set of nonnegative integers and $\lambda$ arises from the use of a real basis\footnote{For $n=0$, the label $\lambda$ becomes spurious.}. The corresponding mode frequencies are given by $\omega_n = 2\pi n/L_o$. The functions $Y^{lm}$, on the other hand, are real scalar harmonics on the sphere, obtained from the real and imaginary parts of the standard complex harmonics with the same value of $m$ \cite{MM}. They provide polar modes. Recall also that $l\in\mathbb{N}_0$ and $m\in\{-l,\cdots,l\}$. Finally, the functions $X^{lm}_{A}$ and $X^{lm}_{AB}$ correspond to the axial components of the real vector and tensor harmonics, while $Z^{lm}_{A}$, $Z^{lm}_{AB}$ and $Y^{lm}_{AB}$ represent the polar components of the real vector, and tensor harmonics \cite{MM}, respectively, where $A$ and $B$ are indices on the sphere\footnote{See e.g. Refs. \cite{DB2,DB3} for the definition of all these harmonics.}. The coefficients of this mode expansion\footnote{Previous studies \cite{MM,MM2} have shown that only modes with $l\geq2$ are physically relevant. Lower multipoles can be discarded: the $l=1$ and $l=0$ (with $n\neq0$) modes correspond to pure gauge degrees of freedom, while the case $n=l=m=0$, i.e the zero-mode, is already accounted for in the background dynamics.} serve as Lagrange multipliers or fundamental variables in phase space for the canonical analysis, and are labeled by $\mathfrak{n} = (n,l,m)$ and $\lambda$. For consistency with earlier works and in consonance with our 3+1 decomposition of the metric, the background dependence is expressed in terms of the interior variables; the exterior counterpart follows directly from the extended ranges obtained for these variables with the transformation \eqref{eq: II-2.5}. Although some intermediate expressions may appear complex, all quantities relevant to the physical discussion remain real\footnote{This subtlety primarily concerns densitized objects, such as the momenta. Negative contributions under square roots originate from density signs that switch when the evolution is radial rather than temporal, ensuring that no imaginary terms arise.}.

Regarding the dynamics, for each mode there are four perturbative gauge constraints (one axial and three polar), which generate first-order perturbative gauge transformations. There is also a perturbative Hamiltonian formed by individual contributions from the modes. The gauge constraints arise from the inherent redundancy of perturbation theory when identifying points between the background and the perturbed spacetimes \cite{DC,MM}. The perturbative Hamiltonian is a quadratic contribution of the first-order perturbations (and therefore a second-order term in the action) which, together with the background Hamiltonian, forms the zero-mode of the total Hamiltonian constraint. It is only this constraint that vanishes, and not necessarily its perturbative part (and even less each of its mode contributions). Collecting all these elements together with the symplectic structure for the perturbative degrees of freedom yields the following expression for the leading-order perturbative contribution to the action:
\begin{equation}
    \label{eq: II-2.10}
    \frac{1}{2}\Delta^2_1[S] = \int\bigg(\sum_{\mathfrak{n},\lambda}\sum_{i=1}^6 \frac{1}{\kappa}p_i^{\mathfrak{n},\lambda}\text{d}h_i^{\mathfrak{n},\lambda} - \bigg[\sum_{i=0}^3C_i[k_i^{\mathfrak{n},\lambda}] + \tilde{H}[\tilde{N}]\bigg]\text{d}\chi\bigg),
\end{equation}
where $C_i$ are the perturbative constraints, with $k_i^{\mathfrak{n},\lambda}$ being their perturbative Lagrange multipliers, and $\tilde{H}$ represents the perturbative Hamiltonian (given by a sum of individual mode Hamiltonians). The notation used emphasizes that this action is a second-order perturbative contribution, entirely composed of first-order products, with no genuine second-order perturbations included. At this truncation order, the axial and polar perturbations remain decoupled, and the description naturally splits into separate contributions for each sector \cite{DC}. To keep the discussion concise, we do not present the full expressions for the perturbative constraints and the Hamiltonian here; instead, we refer the reader to Appendix \ref{app: A}. Together with the background action in Eq. \eqref{eq: II-2.5}, this formulation provides a complete Hamiltonian description of linear BH perturbations, with a phase space that includes background and perturbative degrees of freedom, and dynamics governed by constraints, whether purely perturbative or involving couplings between the perturbations and the background.

A straightforward computation of the number of degrees of freedom shows that, for each mode, the perturbed metric and its conjugate momentum each contain six components (two axial and four polar). The lapse function contributes one polar coefficient, while the shift vector contributes three components (one axial and two polar). Altogether, these add up to ten degrees of freedom (three axial and seven polar), allowing for the reconstruction of the full linear perturbation of the spacetime metric. To make this correspondence explicit, we compare with the first-order Schwarzschild metric perturbation, $\tilde{h} = \sum_{lm} (\tilde{h}^{lm}_{\text{ax}} + \tilde{h}^{lm}_{\text{po}})$, where
\begin{equation}
    \label{eq: II-2.11}
    \begin{aligned}
        &\tilde{h}^{lm}_{\text{ax}} = 2(\tilde{h}_{\zeta}^{lm}X^{lm}_A\text{d}\zeta + \tilde{h}_{\chi}^{lm}X^{lm}_A\text{d}\chi)\text{d}x^A + \tilde{h}_2^{lm}X^{lm}_{AB}\text{d}x^A\text{d}x^B,\\
        &\tilde{h}^{lm}_{\text{po}} = (\tilde{h}_{\zeta\zeta}^{lm}\text{d}\zeta^2 + 2\tilde{h}_{\zeta\chi}^{lm}\text{d}\zeta\text{d}\chi + \tilde{h}_{\chi\chi}^{lm}\text{d}\chi^2)Y^{lm} + 2(j_{\zeta}^{lm}\text{d}\zeta + j_{\chi}^{lm}\text{d}\chi)Z^{lm}_A\text{d}x^A + \chi^2(K^{lm}Y^{lm}_{AB} + G^{lm}Z^{lm}_{AB})\text{d}x^A\text{d}x^B.
    \end{aligned}
\end{equation}
Apart from our specific notation for the \textit{time} and \textit{radial} variables, we otherwise follow a fairly standard convention. Up to a summation over Fourier modes and a rearrangement through simple complex linear combinations (if a basis of complex modes and complex coefficients for them are preferred), the following identifications can be made:
\begin{equation}
    \label{eq: II-2.12}
    \tilde{h}_{\chi}^{lm} = -\kappa k_0^{lm}, \qquad  \tilde{h}_{\zeta}^{lm} = -h_1^{lm}, \qquad \tilde{h}_2^{lm} = h_2^{lm},
\end{equation}
for the axial contributions and, analogously, for the polar ones, 
\begin{equation}
    \label{eq: II-2.13}
    \begin{aligned}
        \tilde{h}_{\zeta\zeta}^{lm} = h_6^{lm}, \qquad \tilde{h}_{\zeta\chi}^{lm} = \kappa \frac{p_b^2}{L_o^2}k_2^{lm}, \qquad \sqrt{-f}\tilde{h}_{\chi\chi}^{lm} = -\kappa\sqrt{\frac{p_b^2}{L_o^2}|p_c|}\tilde{N}k_3^{lm},\\
        j_{\zeta}^{lm} = h_5^{lm}, \qquad j_{\chi}^{lm} = \kappa |p_c|k_1^{lm}, \qquad \chi^2K^{lm} = h_3^{lm}, \qquad \chi^2G^{lm} = h_4^{lm}. 
    \end{aligned}   
\end{equation}
Some of these relations are less straightforward than others, in particular, the $\chi\chi$-component requires careful handling during reconstruction. Since most BH results are obtained in the exterior region, when relating our Hamiltonian formulation to the above expressions, the corresponding allowed ranges of variation for the basic functions of the background variables must be used.

\section{Perturbative gauge invariants \label{sec: III}}

As expected, not all perturbative components correspond to independent physical degrees of freedom. Only certain combinations (the first-order gauge invariants) encode genuine physical information. In the Hamiltonian framework, these can be identified and separated through a canonical transformation that reorganizes the perturbative gauge constraints as phase-space variables. As explained in Appendix \ref{app: A} and Appendix \ref{app: B}, this requires a previous Abelianization of the polar gauge constraints, which is doable at our perturbative order of truncation of the action. Under the aforementioned transformation, the momenta of four out of the six canonical pairs get identified with the perturbative constraints, while the remaining two pairs, constructed to Poisson-commute with the generators of perturbative gauge transformations, represent the linear gauge invariant degrees of freedom \cite{Lan}. Implementing this canonical transformation, and denoting the new perturbative variables by capital letters, we obtain
\begin{equation}
    \label{eq: III-3.1}
    \{h_i^{\mathfrak{n},\lambda},p_{i'}^{\mathfrak{n}',\lambda'}\}_{\text{P}} = \kappa\delta_{ii'}\delta^{\mathfrak{n}\mathfrak{n}'}\delta^{\lambda\lambda'} \qquad \xrightarrow{\text{Canon. Trans.}} \qquad \{Q_i^{\mathfrak{n},\lambda},P_{i'}^{\mathfrak{n}',\lambda'}\}_{\text{P}} = \kappa\delta_{ii'}\delta^{\mathfrak{n}\mathfrak{n}'}\delta^{\lambda\lambda'},
\end{equation}
where the subscript $\text{P}$ indicates that the Poisson brackets are computed with respect to the perturbative variables, disregarding any dependence on the background. The index $i$ runs from $1$ to $6$, with $i=1,2$ labeling the axial components and the remaining ones corresponding to the polar sector. The explicit form of the transformation is provided in Appendix \ref{app: B}. Compared with previous works \cite{MM,MM2}, this procedure is now implemented as a single canonical transformation, rather than a sequence, and the transformation for the polar sector given here incorporates substantial improvements. In earlier studies, one of the imposed requirements was the absence of dilation functions in denominators, a condition motivated by simplifying the quantum analysis \cite{BH_GAB,GBA}. By relaxing this restriction, a broader class of canonical transformations becomes admissible.

In principle, this canonical transformation must act also on the background. Here, however, we restrict attention to the effect on the perturbations. A complementary transformation of the background variables always exists and can be regarded as a second-order correction on the zero-modes, i.e. a correction to the background variables given by quadratic contributions of the first-order perturbations \cite{MM,LMM}. The explicit formulas for the change of background variables corresponding to a background-dependent canonical transformation of the perturbative variables can be found in Appendix \ref{app: D}. Since in the next sections we are going to discuss additional background-dependent canonical transformations of the perturbations (and the composition of several of such transformations is also canonical), we can directly compute the total correction to our background variables at the end. In the following, in all our intermediate steps and in the final results, we will denote the corrected background phase-space variables with the same notation as the original ones to simplify the notation\footnote{We will however use a different notation for the corrected densitized lapse function, for which we will give explicitly the form of its modifications, since these cannot be computed directly from a general formula.}. 

On the other hand, note that, if we only pay attention to first-order modifications of the metric, the quadratic perturbative corrections to the background zero-modes can be ignored. Nevertheless, we insist that they play an important conceptual role: they ensure the preservation of the canonical symplectic structure and make the treatment effectively equivalent to working with a fixed background at the leading-order perturbative truncation of the action. In fact, it is possible to show that, because the canonical transformation depends on the background variables, the perturbative Hamiltonian acquires corrections that reproduce those appearing in fixed-background analyses, with the only distinction that Poisson brackets replace derivatives of the background variables with respect to the evolution parameter (thereby preserving the off-shell character of the formulation) \cite{LMM}. 

As anticipated, the perturbative constraints become now linear combinations of momentum variables,
\begin{equation}
    \label{eq: III-3.2}
    C_0[\textbf{k}_0^{\mathfrak{n},\lambda}] = \sum_{\mathfrak{n},\lambda} \textbf{k}_0^{\mathfrak{n},\lambda}P_2^{\mathfrak{n},\lambda}, \quad C_1[\textbf{k}_1^{\mathfrak{n},\lambda}] = \sum_{\mathfrak{n},\lambda} \textbf{k}_1^{\mathfrak{n},\lambda}P_5^{\mathfrak{n},\lambda}, \quad C_2[\textbf{k}_2^{\mathfrak{n},\lambda}] = \sum_{\mathfrak{n},\lambda} \textbf{k}_2^{\mathfrak{n},\lambda}P_6^{\mathfrak{n},\lambda}, \quad\textbf{C}_3[\textbf{k}_3^{\mathfrak{n},\lambda}] = \sum_{\mathfrak{n},\lambda} \textbf{k}_3^{\mathfrak{n},\lambda}P_4^{\mathfrak{n},\lambda},
\end{equation}
where $\textbf{k}_0^{\mathfrak{n},\lambda}$, $\textbf{k}_1^{\mathfrak{n},\lambda}$, $\textbf{k}_2^{\mathfrak{n},\lambda}$, and $\textbf{k}_3^{\mathfrak{n},\lambda}$ denote the original perturbative Lagrange multipliers, modified with suitable correction terms, as explained in Appendix \ref{app: B}. The change in notation in the last constraint reflects a minor redefinition introduced to obtain commuting perturbative gauge constraints, in accordance with our previous comments. After incorporating also analogous corrections to the lapse function (which are also omitted here for brevity), the Hamiltonian takes the form
\begin{equation}
    \label{eq: III-3.3}
    \kappa\mathbf{\tilde{H}}[\tilde{N}] =  \sum_{\mathfrak{n},\lambda}\frac{\tilde{N}}{2}\left(A_{\text{ax}}[Q_1^{\mathfrak{n},\lambda}]^2 + B_{\text{ax}}[P_1^{\mathfrak{n},\lambda}]^2 + C_{\text{ax}}Q_1^{\mathfrak{n},\lambda}P_1^{\mathfrak{n},\lambda} + A_{\text{po}}[Q_3^{\mathfrak{n},\lambda}]^2 + B_{\text{po}}[P_3^{\mathfrak{n},\lambda}]^2 + C_{\text{po}}Q_3^{\mathfrak{n},\lambda}P_3^{\mathfrak{n},\lambda}\right),
\end{equation}
where all expressions required for this discussion, including the corrections to the Lagrange multipliers, the redefinition of the third constraint and perturbative Hamiltonian, and the new Hamiltonian coefficients, are collected in Appendix \ref{app: B}. The resulting expression naturally splits into two independent contributions. From the structure of the constraints, one identifies $(Q_1^{\mathfrak{n},\lambda},P_1^{\mathfrak{n},\lambda})$ and $(Q_3^{\mathfrak{n},\lambda},P_3^{\mathfrak{n},\lambda})$ as the axial and polar first-order gauge invariant pairs, respectively, offering a natural starting point for constructing the standard master functions used in BH perturbation theory. We do so in the following subsections, where we elaborate on the results of Ref. \cite{MGAC} for the axial sector and present total new results for polar parity.

\subsection{Gerlach-Sengupta gauge invariants \label{ssec: III.A}}

Although a gauge invariant set has already been identified, its dynamical description can still be refined. Since any combination of gauge invariants remains gauge invariant, different, but not independent, collections of variables may describe the same physical content \cite{Lan,LMM}. When such combinations preserve the symplectic structure (i.e., correspond to canonical transformations), they can be employed as phase-space variables that simplify the Hamiltonian. For our purposes, we adopt the following relations for the configuration variables:
\begin{equation}
    \label{eq: III-3.4}
    Q_1^{\mathfrak{n},\lambda} = - \sqrt{\frac{(l+2)!}{(l-2)!}}\mathcal{Q}_1^{\mathfrak{n},\lambda} + 4l(l+1)\frac{L_o^2}{p_b^2}\Omega_b\sqrt{\frac{(l-2)!}{(l+2)!}}\big[\mathcal{P}_1^{\mathfrak{n},\lambda} + 2\Omega_b\mathcal{Q}_1^{\mathfrak{n},\lambda}\big], \qquad Q_3^{\mathfrak{n},\lambda} = \sqrt{\frac{l(l+1)}{(l+2)(l-1)}}\mathcal{Q}_3^{\mathfrak{n},\lambda}.
\end{equation}
From this point, canonical transformations will be restricted to the invariant sector\footnote{See the comments in the previous subsection concerning the corresponding transformation of background variables to maintain a global canonical symplectic structure.}. The conjugate momenta satisfy
\begin{equation}
    \label{eq: III-3.5}
    P_1^{\mathfrak{n},\lambda} = - \sqrt{\frac{(l-2)!}{(l+2)!}}\big[\mathcal{P}_1^{\mathfrak{n},\lambda} + 2\Omega_b\mathcal{Q}_1^{\mathfrak{n},\lambda}\big], \qquad P_3^{\mathfrak{n},\lambda} = \sqrt{\frac{(l+2)(l-1)}{l(l+1)}}\bigg[\mathcal{P}_3^{\mathfrak{n},\lambda} - \frac{C_{\text{po}}}{2}\mathcal{Q}_3^{\mathfrak{n},\lambda}\bigg], 
\end{equation}
where $C_{\text{po}}$ is defined explicitly in Appendix \ref{app: B}. Even though the formalism allows for parity mixing, modes of different parity are not coupled here to preserve all the symmetries of the dynamical equations of the perturbations. Exploring such transformations would constitute an interesting direction for future work.

Thanks to the above transformation, discussed with more detail in Appendix \ref{app: C}, the resulting Hamiltonian adopts a purely quadratic form,
\begin{equation}
    \label{eq: III-3.6}
    \kappa\bar{\mathcal{H}}[\tilde{N}] = \sum_{\mathfrak{n},\lambda} \frac{\tilde{N}}{2}\Big( [\mathcal{P}_1^{\mathfrak{n},\lambda}]^2 + \mathcal{V}^{\text{ax}}[\mathcal{Q}_1^{\mathfrak{n},\lambda}]^2 + [\mathcal{P}_3^{\mathfrak{n},\lambda}]^2 + \mathcal{V}^{\text{po}}[\mathcal{Q}_3^{\mathfrak{n},\lambda}]^2 \Big),
\end{equation}
where the potentials, encoding the relevant physical content of the perturbations, are written, using a compact notation, in terms of the background functions as
\begin{equation}
    \label{eq: III-3.7}
    \mathcal{V}^{\text{ax}/\text{po}} = (\omega_n^2 - V_l^{\text{ax}/\text{po}})p_c^2 - \bigg(\Omega_b^2+\frac{p_b^2}{L_o^2}\bigg).
\end{equation}
The Fourier frequency is isolated, up to an overall factor of $p_c^2$. The remaining terms, aside from those that dynamically compensate for $p_c^2$, can then be interpreted as a generalization of the Regge-Wheeler and Zerilli potentials, 
\begin{equation}
    \label{eq: III-3.8}
    \begin{aligned}
        &V_l^{\text{ax}} = -\frac{1}{p_c^2}\bigg[l(l+1)\frac{p_b^2}{L_o^2} + 6\Omega_b\Omega_c\bigg],\\
        &V_l^{\text{po}} = -\frac{1}{p_c^2}\bigg[(l+2)^2(l-1)^2\bigg(l(l+1)\frac{p_b^2}{L_o^2} - 6\Omega_b\Omega_c\bigg) + \bigg(6\Omega_b\Omega_c\frac{L_o^2}{p_b^2}\bigg)^2\bigg((l+2)(l-1)\frac{p_b^2}{L_o^2} - 2\Omega_b\Omega_c\bigg)\bigg]\frac{1}{\Lambda^2},
    \end{aligned}
\end{equation}
with $\Lambda = (l+2)(l-1) - 6\Omega_b\Omega_cL_o^2/p_b^2$, and where the subindex $l$ emphasizes the dependence on the angular momentum number. 

This setup provides a more unified formulation compared to previous works \cite{MM,MM2}. Classically, the equations of motion, up to simple combinations, lead to wave-like equations for the configuration variables governed by the four-dimensional Laplace-Beltrami operator. While these pairs do not yet have a direct correspondence with the BH master functions, they already constitute interesting results on their own. For instance, at the quantum level, they enable a more direct discussion of the uniqueness of the Fock representation in the polar sector since the Fourier frequency is properly isolated, allowing for an analysis closer to that of the axial counterpart \cite{MM}.

\subsection{Hamiltonian master functions \label{ssec: III.B}}

The variables introduced above already constitute a suitable canonical set. However, one additional step is needed to establish a clean connection with the most usual BH master functions. We therefore perform the following canonical transformation:
\begin{equation}
    \label{eq: III-3.9}
    \tilde{\mathcal{Q}}_j^{\mathfrak{n},\lambda} = \sqrt{|p_c|}\mathcal{Q}_j^{\mathfrak{n},\lambda}, \quad \mathcal{P}_j^{\mathfrak{n},\lambda} = \sqrt{|p_c|}\tilde{\mathcal{P}}_j^{\mathfrak{n},\lambda} - \Omega_b\mathcal{Q}_j^{\mathfrak{n},\lambda}, \quad \text{for} \quad j=1,3,
\end{equation}
which removes the $p_c^2$ factor from the Hamiltonian term with the squared Fourier frequency. After this transformation, the Hamiltonian,
\begin{equation}
    \label{eq: III-3.10}
    \kappa\tilde{\mathcal{H}}[\tilde{N}] = \sum_{\mathfrak{n},\lambda} \frac{\tilde{N}}{2}|p_c|\Big([\tilde{\mathcal{P}}_1^{\mathfrak{n},\lambda}]^2 + (\omega_n^2 - V_l^{\text{ax}})[\tilde{\mathcal{Q}}_1^{\mathfrak{n},\lambda}]^2 + [\tilde{\mathcal{P}}_3^{\mathfrak{n},\lambda}]^2 + (\omega_n^2 - V_l^{\text{po}})[\tilde{\mathcal{Q}}_3^{\mathfrak{n},\lambda}]^2\Big),
\end{equation}
has the frequency contribution isolated. Aside from the densitized lapse, however, a global factor of $|p_c|$ remains, altering the natural choice of evolution parameter. In direct analogy with the use of tortoise coordinates in the Schwarzschild case, we introduce the reparametrization $\text{d}\eta = |p_c|\tilde{N}\text{d}\chi$ (indeed, for the Schwarzschild solution we recover $\text{d}\eta = \text{d}\chi/f$). 

After performing a Fourier summation to highlight the importance of the two-dimensional set of spherical orbits, the equations of motion for the configuration fields obtained in those two dimensions (each of these functions with associated harmonic labels $l$ and $m$) take the form
\begin{equation}
    \label{eq: III-3.11}
    \big[ -\partial_{\zeta}^2 + \partial_{\eta}^2 - V_l^{\text{ax}/\text{po}} \big]\tilde{\mathcal{Q}}_j^{lm} = 0, \quad \text{for} \quad j=1,3,
\end{equation}
which exhibit a wave-like structure. This yields a generalized expression for the master equations that is valid in the interior and exterior regions. Up to a global factor, the expression can also be written in terms of the two-dimensional Laplace-Beltrami operator on this submanifold.

Particularizing the background to the Schwarzschild solution (for instance, using the interior expressions of Eq. \eqref{eq: II-2.6}, with quadratic corrections of the perturbations ignored) allows us to recover then the BH master equations with the standard Regge-Wheeler and Zerilli potentials,
\begin{equation}
    \label{eq: III-3.12}
    V_{l,\text{Sch}}^{\text{ax}} = f\bigg[\frac{l(l+1)}{\chi^2} -\frac{ 6M}{\chi^3}\bigg],\quad V_{l,\text{Sch}}^{\text{po}} = \frac{f}{\Lambda_{\text{Sch}}^2}\bigg[\frac{(l+2)^2(l-1)^2}{\chi^2}\bigg(l(l+1) + \frac{6M}{\chi}\bigg) + \frac{36 M^2}{\chi^4}\bigg((l+2)(l-1) + \frac{2M}{\chi}\bigg)\bigg],
\end{equation}
with $\Lambda_{\text{Sch}} = (l+2)(l-1) + 6M/\chi$, and the subindex ${\text{Sch}}$ indicating evaluation in the Schwarzschild case. This partial result identifies the master equations, but not yet the corresponding master functions. To determine the latter, we rewrite the invariants in terms of the original canonical variables and, upon imposing the Schwarzschild background, compare them with the standard expressions in the literature \cite{BCS}. Avoiding for convenience the use of the overbar notation when working with the exterior background variables, as discussed in Subsec. \ref{ssec: II.A}, and tracing back our canonical transformations, we first obtain
\begin{equation}
    \label{eq: III-3.13}
    \begin{aligned}
        &\tilde{\mathcal{Q}}_1^{\mathfrak{n},\lambda} = \sqrt{\frac{(l-2)!}{(l+2)!}}\sqrt{|p_c|}\bigg[p_1^{\mathfrak{n},\lambda} + 4l(l+1)\frac{L_o^2}{p_b^2}\Omega_b h_1^{\mathfrak{n},\lambda}\bigg],\\ 
        &\tilde{Q}_3^{\mathfrak{n},\lambda} = \sqrt{\frac{l(l+1)}{(l+2)(l-1)}}\sqrt{|p_c|}\bigg[\frac{h_3^{\mathfrak{n},\lambda}}{|p_c|} + \frac{l(l+1)}{2}\frac{h_4^{\mathfrak{n},\lambda}}{|p_c|} + \frac{2}{\bar{\Lambda}}\frac{L_o^2}{p_b^2}\frac{\Omega_b}{|p_c|}\bigg(\frac{L_o^2}{p_b^2}\bar{p}_c^2\Omega_bh_6^{\mathfrak{n},\lambda} -\frac{p_b^2}{L_o^2}p_6^{\mathfrak{n},\lambda} + 2\Omega_bh_3^{\mathfrak{n},\lambda}\bigg)\bigg].
    \end{aligned}
\end{equation}
The perturbative momenta $p_1^{\mathfrak{n},\lambda}$ and $p_6^{\mathfrak{n},\lambda}$ must be reformulated in terms of the metric perturbations and their derivatives. This is achieved by employing the perturbative equations of motion\footnote{Imposing these equations places the perturbations on shell, though the background is still treated off shell.}, which follow from Eq. \eqref{eq: II-2.10} and the expressions in Appendix \ref{app: A} adapted to the exterior background.

After substituting these relations, summing over Fourier modes, and including the following convenient prefactors for the CPM and ZM master functions:
\begin{equation}
    \label{eq: III-3.14}
     \Psi_{\text{CPM}}^{lm} = 2\sqrt{\frac{(l-2)!}{(l+2)!}}\tilde{\mathcal{Q}}_1^{lm}, \qquad \Psi_{\text{ZM}}^{lm} = \sqrt{\frac{(l+2)(l-1)}{l(l+1)}}\frac{2}{l(l+1)}\tilde{\mathcal{Q}}_3^{lm},
\end{equation}
we obtain for these functions a generalized form in terms of the original metric, lapse, and shift perturbations, introduced in Eq. \eqref{eq: II-2.9}. They read as
\begin{equation}
    \label{eq: III-3.15}
    \begin{aligned}
        &\Psi_{\text{CPM}}^{lm} = \frac{2}{(l+2)(l-1)}\frac{\sqrt{|p_c|}}{\tilde{N}}\bigg[-\partial_{\chi}h_1^{lm} + \kappa \partial_{\zeta}k_0^{lm} + 2\tilde{N}\Omega_bh_1^{lm}\bigg]\frac{L_o^2}{p_b^2},\\
        &\Psi_{\text{ZM}}^{lm} = \frac{2}{l(l+1)}\sqrt{|p_c|}\bigg[\frac{h_3^{lm}}{|p_c|} + \frac{l(l+1)}{2}\frac{h_4^{lm}}{|p_c|} - \frac{2}{\bar{\Lambda}}\frac{\Omega_b}{|p_c|}\bigg(\Omega_b|p_c|\kappa k_3^{lm} + l(l+1)\frac{|p_c|}{\tilde{N}}\kappa k_1^{lm} + \frac{\partial_{\chi}h_3^{lm}}{\tilde{N}} - 2\Omega_bh_3^{lm}\bigg)\frac{L_o^2}{p_b^2}\bigg].
    \end{aligned}
\end{equation}
By specializing the background to Schwarzschild and using the dictionary, formulated for the exterior background, between our canonical perturbations and the Schwarzschild perturbations derived in Eqs. \eqref{eq: II-2.12} and \eqref{eq: II-2.13}, we finally arrive at the commonly used expressions
\begin{equation}
    \label{eq: III-3.16}
    \begin{aligned}
        &\Psi_{\text{CPM},{\text{Sch}}}^{lm} = \frac{2\chi}{(l+2)(l-1)}\bigg[\partial_{\chi}\tilde{h}_{\zeta}^{lm} - \partial_{\zeta}\tilde{h}_{\chi}^{lm} - \frac{2}{\chi}\tilde{h}_{\zeta}^{lm}\bigg],\\
        &\Psi_{\text{ZM},{\text{Sch}}}^{lm} = \frac{2\chi}{l(l+1)}\bigg[K^{lm} + \frac{l(l+1)}{2}G^{lm} + \frac{2f}{\Lambda_{\text{Sch}}}\bigg( f\tilde{h}_{\chi\chi}^{lm} - \frac{l(l+1)}{\chi}j_{\chi}^{lm} - \chi\partial_{\chi}K^{lm}\bigg)\bigg],
    \end{aligned}
\end{equation}
where we have used the notation explained in Eq. \eqref{eq: II-2.11}.

The pairs $(\tilde{\mathcal{Q}}_1^{\mathfrak{n},\lambda},\tilde{\mathcal{P}}_1^{\mathfrak{n},\lambda})$ and $(\tilde{\mathcal{Q}}_3^{\mathfrak{n},\lambda},\tilde{\mathcal{P}}_3^{\mathfrak{n},\lambda})$ correspond (up to an overall $l$-dependent but otherwise constant factor) to a generalization of the BH master functions and exhibit a particularly simple Hamiltonian structure, with a diagonal form in which the momenta appear with unit coefficients and the configuration variables are multiplied by a potential plus an isolated squared Fourier frequency (apart from a global factor of $|p_c|$ in the perturbative Hamiltonian). The Hamiltonian framework, however, retains some freedom, allowing the use of alternative canonical pairs that preserve the key properties of the BH master functions. Indeed, there exists an entire family of such pairs, related by canonical transformations that maintain this Hamiltonian structure but differ in their associated potentials. This feature, well known in the BH perturbation literature as the Darboux symmetry, emerges naturally from the canonical requirements themselves \cite{MGAC}.

\section{Hybrid quantization \label{sec: IV}} 

So far, we have identified the gauge invariant combinations associated with the BH master functions. In our Hamiltonian framework, these invariants appear as configuration variables in a phase space that dynamically incorporates the BH background and its perturbations\footnote{As we have noticed before, and although not immediately evident, preserving a global symplectic structure with a dynamical background in the perturbative description requires the construction that is detailed in the four appendices. In this construction, perturbative corrections are introduced to the Lagrange multipliers and to the background variables \cite{LMM}. Recall that, for the corrected background variables, we use the same notation as for the original ones, to simplify the expressions.}. The equations of motion, which reduce to the master equations for the perturbations, follow from a total Hamiltonian constraint (already integrated over the whole three-dimensional section) which couples the perturbative and nonperturbative sectors. At the truncation order considered in the action, this leads to a simple perturbative formulation incorporating backreaction effects on the background and providing a natural starting point for the canonical quantization of the system.

In this work, we adopt a hybrid quantization scheme that allows for a simultaneous, and potentially distinct, treatment of the background and the perturbations \cite{hyb-review}. The hybrid-LQC approach is grounded in the idea that, as a good approximation to a full quantum gravity theory, systems with perturbations should admit a regime in which quantum field theory in curved spacetime remains reliable. In this setting, the zero-modes describing the background are expected to capture the leading quantum gravity effects, while the (nonzero-mode) perturbations can be described with more conventional techniques. The framework is sufficiently general, with only some consistency restrictions on the specific quantum representations. The most relevant restrictions arise from the construction of the quantum operator for the total Hamiltonian constraint, in particular the need for a proper definition of all the quantum background operators entering its definition. This typically involves inverse geometric operators that are generically ill-defined in standard representations for the background at the counterpart of classical singularities, as it is the case in geometrodynamics. But, although originally developed in the context of LQC \cite{FMO,hyb1,hyb-review}, the hybrid approach is not limited to it and can be combined with other quantum gravity formalisms for the background as well.

As an extension of earlier works \cite{MM,MM2}, here we continue to employ an LQC representation for the background and a Fock quantization for the perturbative sector. To avoid unnecessary repetition, we summarize only the key elements of both kinematic representations, focusing instead on the new features introduced by our choice of canonical variables and on the representation of the Hamiltonian constraint with the new potentials.

\subsection{Kinematic Hilbert space \label{ssec: IV.A}}

The phase space sector described by the (corrected) background variables is represented using the discrete framework of LQC \cite{BH_GAB,GBA}. Since the background variables are related to densitized triads and connections, in this framework one constructs an holonomy-flux algebra and determines a representation for it in which fluxes, describing triads, have a discrete spectrum \cite{AP}. Our choice of this quantum framework for the background is motivated by its success in the quantization of other cosmological backgrounds (Friedmann-Lemaître-Robertson-Walker, Bianchi I, etc.) \cite{IJV}, by the fact that our background can actually be regarded as a cosmological model in the interior region, and by the good physical properties that this quantum treatment promotes, such as the resolution of strong singularities \cite{APS1,APS2}. From a purely kinematic perspective, where the aim is to construct an adequate Hilbert space to represent our relevant phase-space functions, the distinction between interior and exterior variables is immaterial, since their differences appear only in the ranges allowed for them classically on shell, and at the dynamical level in the interpretation of the Hamiltonian evolution. 

The construction followed in previous works \cite{BH_GAB,GBA} requires extending the background phase space to include two regularization parameters  of quantum origin, $\delta_b$ and $\delta_c$, as additional canonical variables. These parameters implement conditions on the smallest holonomy circuits that can be constructed in the quantum theory on the surface that replaces the BH singularity \cite{AOS2}. To compensate for this enlargement, two extra constraints, which encapsulate the aforementioned conditions, are introduced to control those parameters \cite{BH_GAB}. The correct classical description is recovered in the limit in which the parameters are treated as negligible. Conversely, one may adopt an effective description in which quantum effects are incorporated by directly replacing the background connection variables with classical holonomies along the circuits determined by the introduced parameters. These holonomies are denoted by $\mathcal{N}_{\mu_b} = \exp\{ib\mu_b/2\}$ and $\mathcal{N}_{\mu_c} = \exp\{ic\mu_c/2\}$, where $\mu_b$ and $\mu_c$ are two real-valued parameters determining the coordinate length associated with the considered edges. The conjugate triad variables are then regarded as fluxes \cite{AOS,BH_GAB}. We refer to the analysis in Ref. \cite{GBA} for further details.

As we have anticipated, the quantization of this extended background phase space employs an LQC representation for the geometric degrees of freedom and, for simplicity, a Schrödinger representation for the regularization parameters. In the geometric subsector, the kinematic Hilbert space is constructed in the triad representation by introducing two copies of the holonomy-flux algebra, one for the $b$--sector and another for the $c$--sector, completing each of them with respect to the discrete inner product, and then taking the tensor product of the two resulting Hilbert spaces \cite{BH_GAB,GBA}. In contrast, the subsector associated with the regularization parameters is quantized by taking the tensor product of two copies (one for each parameter) of the Hilbert space of square-integrable functions on the real line equipped with the Lebesgue measure. The full extended kinematic Hilbert space is obtained as the tensor product of the Hilbert spaces of these two subsectors.

With this construction in place, and after a few minor redefinitions that render the action of the basic holonomy functions independent of the regularization parameters (namely, a rescaling so that the fluxes absorb the regularization parameters as $\tilde{p}_b = p_b/\delta_b$ and $\tilde{p}_c = p_c/\delta_c$) every background-dependent quantity, including the dilation functions and positive or negative powers of the triads, admits a well-defined quantum counterpart. In the geometric sector, the action of the basic holonomy-flux elements is given by
\begin{equation}
    \label{eq: IV-4.1}
    \hat{\mathcal{N}}_{\delta_b} |\tilde{\mu}_b \rangle = |\tilde{\mu}_b + 1 \rangle, \qquad \hat{\mathcal{N}}_{\delta_c} |\tilde{\mu}_c \rangle = |\tilde{\mu}_c + 1 \rangle, \qquad \hat{\tilde{p}}_b |\tilde{\mu}_b\rangle = \frac{\gamma \tilde{\mu}_b}{2} |\tilde{\mu}_b\rangle, \qquad \hat{\tilde{p}}_c |\tilde{\mu}_c\rangle = \gamma \tilde{\mu}_c  |\tilde{\mu}_c\rangle,
\end{equation}
where $|\tilde{\mu}_b,\tilde{\mu}_c\rangle = |\tilde{\mu}_b\rangle\otimes|\tilde{\mu}_c\rangle$ denotes the triad-eigenstate basis of the geometric kinematic Hilbert space. A key ingredient enabling the representation of the dilation functions is the MMO prescription (named after Martín-Benito, Mena Marugán, and Olmedo) \cite{MMO}, which provides a consistent treatment of the noncommuting factors appearing in the dilation operators. Explicitly,
\begin{equation}
    \label{eq: IV-4.2}
    \hat{\Omega}_b = \frac{1}{2\gamma L_o}|\hat{\tilde{p}}_b|^{1/2}\left[\widehat{\sin(\delta_b b)}\widehat{\text{sign}(\tilde{p}_b)}+\widehat{\text{sign}(\tilde{p}_b)}\widehat{\sin(\delta_b b)}\right]|\hat{\tilde{p}}_b|^{1/2}, \qquad \widehat{\sin{(\delta_ b b)}} = \frac{1}{2i} (\hat{\mathcal{N}}_{2\delta_b}-\hat{\mathcal{N}}_{-2\delta_b}), 
\end{equation}
with a completely similar definition of $\hat{\Omega}_c$ with the interchange of the labels $b$ and $c$, and
where $\widehat{\text{sign}}$ is the sign operator. With this prescription, the holonomies and the signs of the triad variables $p_b$ and $p_c$ are symmetrized \`a la Weyl, while the result is symmetrized algebraically with any power of the norm of those triad variables. The corresponding action of a dilation function can then be viewed as a second-order difference operator with constant displacement in the triad labels. 

This action leaves invariant only certain superselected subspaces, consisting of states with support on semilattices for the triad variables $p_b$ and $p_c$ (each of them suitably scaled by one of the regularization parameters of the model) \cite{GBA}. Since the background Hamiltonian depends only on these triad variables and on the dilation functions, its analysis (and therefore that of the constrained quantum background) can be restricted to a single separable superselected Hilbert subspace. In any such sector, the allowed triad eigenvalues become numerable and with definite sign (e.g. strictly positive), and therefore never reach zero, ensuring that their square roots and inverses are well defined. This is an important property, later required for a proper definition of the perturbative contribution to the total Hamiltonian of the perturbed system, as we will see. Finally, let us mention that the dilation operators can be shown to have a purely continuous and nondegenerate spectrum for each of the superselection sectors, equal to the real line \cite{GBA}.
  
The pairs of gauge invariants corresponding to the master functions studied in the previous section are represented using a Fock quantization, a natural choice for recovering quantum field theory in curved spacetime \cite{hyb-review}. The focus on the gauge invariant sector is well justified: all first-order perturbative constraints are of momentum type and mutually commuting under Poisson brackets, so they can be represented in terms of commuting quantum operators that act as generalized (functional) derivatives. Imposing that these operators annihilate physical states, the quantum version of the classical constraints guarantees that such quantum states are independent of the perturbative gauge degrees of freedom. On the other hand, an adaptation of the analysis presented in Ref. \cite{AT} proves that we can select a family of unitarily equivalent Fock representations for the gauge invariant sector by demanding invariance of the vacuum under background isometries and the unitary implementability of the Heisenberg dynamics when the background is treated as classical \cite{AT}. 

Let us be more specific about the proof of uniqueness of this family of representations. For both the axial and the polar sectors, the proof relies on the existence of a nontrivial dynamics for the creation and annihilation operators that is unitarily implementable, once the classical background isometries have been imposed. In our setting, the proof follows very similar arguments to those explained for a perturbative scalar field in Ref. \cite{AT}, which in turn are based on previous results about perturbations of anisotropic cosmologies achieved in Ref. \cite{BIInf}. The only subtlety compared with Ref. \cite{AT} is the presence of a different global factor multiplying the squared Fourier frequency in the Hamiltonian \eqref{eq: III-3.10}, leading to a distinct choice of evolution parameter. Then, the procedure employed in Ref. \cite{AT} allows to isolate in the Hamiltonian a background-independent contribution proportional to $k^2 = \omega_n^2 + l(l+1)$, which can be interpreted as a constant (angular) wave number on the space of mode labels $(n,l)$. Added to this quantity, we obtain new terms that can be interpreted as a background-dependent mass such that its limit when $k$ tends to infinity is finite, analogous to the situation found in Refs. \cite{MM,MM2,AT}. More concretely, this mass depends on simple background-dependent functions of the unit vector $\hat{l}$, defined as $\hat{l}^2 = l(l+1)/k^2$, but is otherwise independent of $k$ except for subdominant contributions of order $O(k^{-2})$. Actually, subdominant contributions of this type do not arise in the case discussed in Ref. \cite{AT}, but they already appeared in the analysis of polar perturbations carried out in Ref. \cite{MM2}, where it was shown that they do not affect the uniqueness of the Fock representation.

These considerations motivate the introduction of a preferred set of creation and annihilation-like variables for each perturbative mode of the invariant sector that can be viewed as those corresponding to the case of vanishing mass for modes with wave number $k$, which is a simple election among the unitarily equivalent Fock representations satisfying our requirements of symmetry invariance and unitary Heisenberg evolution \cite{AT}. At this point of the discussion, it is worth noting that, in the case with compact sections that we are considering, the representations corresponding to any constant value of the mass are unitarily equivalent\footnote{Inequivalence arises only in the noncompact limit, owing exclusively to infrared issues, which can be ignored in practice if the relevant scales for the physical analysis are finite.}. In this sense, it is remarkable that, if we restrict to modes with angular momentum label $l$ bounded from above (so that only a finite number of values of $l$ are possible), we can equivalently adopt the representation that would be naturally associated with the massless case for modes with wave number equal to the Fourier frequency $\omega_n$.  

The resulting Fock space is then constructed in the standard manner, with a basis given by occupation-number states specified by a multi-index of mode occupation numbers with only finitely many nonzero entries.

\subsection{Total Hamiltonian constraint operator \label{ssec: IV.B}}

The tensor product between the background kinematic Hilbert space and the Fock space of the gauge invariant perturbations provides the representation space in which, in principle, we have to implement the remaining constraint, the total Hamiltonian constraint, at the quantum level. Since this constraint couples background and perturbative degrees of freedom, its quantization is less direct than that of the purely perturbative constraints. The quantum constraint plays a fundamental role, because states annihilated by its action provide the physical states of the system. 

Hence, let us focus our attention on obtaining a suitable operator representation of the constraint, which will later allow us to construct physical solutions. Classically, the constraint is given by $\tilde{H}_{\text{T}} = \tilde{H}_{\text{KS}} + \tilde{\mathcal{H}}$. The explicit expressions of its two contributions can be found in Eqs. \eqref{eq: II-2.3} and \eqref{eq: III-3.10}. Because we have kept a consistent density weight in both the background and the perturbative sectors, these two contributions combine directly (up to irrelevant global factors) into the relation
\begin{equation}
    \label{eq: IV-4.3}
     \Omega_b^2 + \frac{p_b^2}{L_o^2} + 2\Omega_b\Omega_c - \frac{|p_c|}{\kappa L_o}  \sum_{\mathfrak{n},\lambda} \Big([\tilde{\mathcal{P}}_1^{\mathfrak{n},\lambda}]^2 + (\omega_n^2 - V_l^{\text{ax}})[\tilde{\mathcal{Q}}_1^{\mathfrak{n},\lambda}]^2 + [\tilde{\mathcal{P}}_3^{\mathfrak{n},\lambda}]^2 + (\omega_n^2 - V_l^{\text{po}})[\tilde{\mathcal{Q}}_3^{\mathfrak{n},\lambda}]^2\Big) = 0,
\end{equation}
which serves as the starting point for the quantum theory. For clarity, we recall that the background variables already encode the perturbative corrections arising from our canonical construction, although we keep for them the original notation. 

When promoting Eq. \eqref{eq: IV-4.2} to an operator, the nontrivial elements are the combinations involving the axial and polar potentials, to which we now turn. A key simplification in our analysis comes from a well-known result in the literature: the Chandrasekhar transformation, which relates the polar and axial potentials \cite{Chandra}. This correspondence (together with the associated correction to the background variables, assumed to be implemented again without changing our notation) allows us to focus exclusively on the Regge-Wheeler potential and leave for future work the investigation of possible quantum differences that may arise when the Zerilli potential is treated directly. Therefore, in practice, the criteria we employ for the axial sector extend straightforwardly to its polar counterpart. We thus restrict attention to the axial perturbations for simplicity.

Although the axial potential is compactly written in Eq. \eqref{eq: III-3.8}, a minor reformulation is convenient for our purposes. The reason is tied to the notion of evolution that we wish to employ. In a fully covariant system such as the one under consideration, a meaningful quantum evolution requires selecting an internal variable with respect to which physical quantities can be compared. Among several possible choices, we adopt $\Omega_b$ as our evolution variable. This choice seems reasonable, because $\Omega_b$ has classically a positive Hamiltonian derivative (given by $2\gamma p_b^2/L_o^2$), and plays a clear role in the classical Schwarzschild solution, providing a suitable geometric reference for the evolution \cite{BAl}.

With this motivation, it is useful to rewrite the Regge-Wheeler potential in terms of $\Omega_b$ using the background Hamiltonian, which vanishes up to quadratic perturbative contributions owing to the total Hamiltonian constraint, and hence at all practical purposes in our order of truncation for the perturbative Hamiltonian. Thus, we get the following equivalent expression for the potential:
\begin{equation}
    \label{eq: IV-4.4}
    \tilde{V}_l^{\text{ax}} = \frac{1}{p_c^2}\left(l(l+1)\Omega_b^2 + 2[l(l+1)-3]\Omega_b\Omega_c\right),
\end{equation}
which does not explicit depend on $p_b^2$. Since $\Omega_c$ is a constant of motion, this form naturally suggests a change of representation (or \textit{picture}) in the $b$--sector of the background Hilbert space. Instead of working in the original triad basis, we may equivalently adopt a representation diagonalizing the operator associated to $\Omega_b$. Since we know that this dilation operator is essentially self-adjoint in each superselection sector, with continuous spectrum over the entire real line \cite{GBA}, the desired change of representation is implemented through the corresponding spectral decomposition of the identity. Its generalized eigenfunctions define a unitary transformation from the triad representation to the $\Omega_b$--representation, where the $\Omega_b$--operator acts multiplicatively. The $c$--sector, in contrast, remains in the standard triad representation, which (as discussed previously) poses no difficulties regarding the definition of the inverse operator for $p_c$. In fact, this even allows us to introduce well-defined operators for any negative algebraic power of $|p_c|$, a result that will be useful in a moment. In summary, expressing the Regge-Wheeler potential entirely in terms of $\Omega_b$ considerably simplifies the quantization of the background dependence in the perturbative Hamiltonian. Unlike in the perturbative sector, in the homogeneous background the dependence on $p_b^2$ cannot be removed. When changing representation, one must therefore handle the action of this operator with care, as it no longer acts by simple multiplication in the new basis.

A second technical issue arises from the appearance of operator products involving $p_c$ and $\Omega_c$, which do not commute. The most direct solution is to adopt an algebraic symmetrization prescription in the norm of $p_c$, something that is feasible thanks to our comments above. Concretely, since $p_c$ acts multiplicatively in our representation, whenever a factor of $\Omega_c$ multiplies a positive function $f(p_c)$ we replace the product in the quantum counterpart by the direct operator version of $\sqrt{f(p_c)}\Omega_c\sqrt{f(p_c)}$, thereby removing ordering ambiguities. With these ordering prescriptions and the reformulation of the potential described above, the promotion of the total Hamiltonian constraint to a quantum operator becomes well defined. Let us conclude with two important remarks. First, note that the perturbative Hamiltonian has been represented in a way that preserves the superselected Hilbert subspaces of the background. Lastly, it is worth emphasizing that the invertibility of the triad variable $p_c$ has been a key property to build a proper quantum representation of the perturbative Hamiltonian.

\section{Conclusion \label{sec: V}}

In this work, we have presented a generalization of the BH master functions in the context of a Hamiltonian formalism for perturbed nonrotating BHs, formulated in terms only of functions of the background geometry, perturbative gauge invariants, perturbative gauge constraints, and perturbative gauge degrees of freedom. The construction applies to the interior and exterior regions (up to an extension in the classically allowed range for certain functions of the background variables), and can be extended to effective models with modified backgrounds, including the case of quantum modifications. For a Schwarzschild solution, the standard invariants (CPM and ZM) are naturally recovered. Dynamically, the model is simple, with a Hamiltonian structure that allows a direct quantum treatment. Using hybrid-LQC, we obtain a unified quantum description of the background and the perturbations, providing a foundational framework that could support future quantum predictions for BHs, including issues related with QNMs \cite{MGAC,BAl}.

Classically, by consolidating several partial results from earlier Hamiltonian approaches to BH perturbations, and after minor changes in notation and a refinement in the treatment of polar variables, we have obtained a concise canonical formulation for both axial (see Ref. \cite{MGAC}) and polar sectors. With a new choice of canonical variables for the perturbative gauge invariants, the master functions become configuration fields, described by mode coefficients that are variables of the perturbative gauge invariant sector of the phase space. In turn, their conjugate momenta are redefined to relate to the master functions through a derivative. Given that most canonical pairs for the metric perturbations reduce to a constraint and a pure perturbative gauge quantity, the physical content of the perturbations is indeed isolated in a subspace described by two canonical pairs for each perturbative mode. The dynamics of the gauge invariants then take a form corresponding to (generalized) harmonic oscillators with potentials reproducing the Regge-Wheeler and Zerilli expressions, with our choices of perturbative configuration variables for each mode in the gauge invariant sector.

A complex canonical transformation of a canonical pair for the background resolves the difficulties of the exterior Hamiltonian formulation. This transformation amounts to extending the range of the squared conformal factor $p_b^2$ of the set of spherically symmetric orbits from the positive to the negative axis, while maintaining real the associated dilation function. The staticity of the geometry no longer obscures the meaning of evolution. Instead of a standard time-like description, we adopt, inspired by the interior formulation, a radial evolution that naturally extends interior results. Other Hamiltonian approaches describing such a radial evolution have also been discussed recently in the literature (see e.g. Refs. \cite{Livine,koc}). Since the aforementioned canonical transformation does not involve any change of coordinates, it provides a unified description with a single evolution parameter for both regions. The transformation also applies directly to the perturbations, as the background dependence of the perturbative Hamiltonian enters only through the dilation function (or alternatively through it and $p_b^2$) in the $b$--sector. Moving from interior to exterior results thus reduces to transforming the background variables and reinterpreting the Hamiltonian evolution.

The quantization of the system supports different schemes for treating background and perturbative degrees of freedom, under the assumption that the background geometry dominates over the perturbations. Choosing LQC to construct the specialized representation for the background allows background functions to be defined as consistent operators, and solutions of the quantum background Hamiltonian constraint can be characterized by their mass profile \cite{GBA}.

The LQC representation can be extended to include the perturbations by combining it with a Fock representation (unique up to unitary equivalence) of the canonical pairs of gauge invariants that describe the modes of the master functions (and their derivatives in the $\zeta$--direction). The simplicity of the formulation naturally enforces gauge invariance of physical states through the quantum constraints and provides a simple expression for the total Hamiltonian constraint, thanks to the Regge-Wheeler potential and the Chandrasekhar transformation that relates the Zerilli potential with it.

Future lines of investigation include the study of effective geometries for BHs within the canonical framework. A deeper theoretical analysis of the master functions is also possible. In the axial sector, recent progress has shown that the well-known Darboux symmetry arises from a specific set of canonical transformations that preserve a particular Hamiltonian structure \cite{MGAC}. Generalizing this last result to include the polar sector is under investigation and seems achievable. On the quantum side, deriving effective equations analogous to the classical master functions for QNMs, emerging from a fully quantum evolution, is also under consideration \cite{BAl}. Note also that our formalism allows to study the backreaction on the background up to second-order in the perturbations. In fact, we have employed corrected zero-mode background variables which already contain quadratic contributions of the perturbations. These corrected variables, on the other hand, must satisfy, together with the perturbative gauge invariants, the total Hamiltonian constraint, at second order in the perturbative hierarchy. Therefore, this Hamiltonian further provides direct information about the backreaction up to the considered truncation order in the perturbations. Finally, a long-term goal is to obtain generalized equations for perturbations in rotating BHs that can be connected with the Teukolsky equations \cite{T}.

\acknowledgments

The authors are very thankful to C.F. Sopuerta for useful discussions. This work was partially supported by MCIN/AEI/10.13019/501100011033 and FSE+ under the Grant No. PID2023-149018NB-C41. A. M.-S. acknowledges support from the PIPF-2023 fellowship from Comunidad Aut\'onoma de Madrid. The reference number is PIPF-2023/TEC-30167.

\newpage

\appendix

\section{Initial perturbative expressions \label{app: A}}

In this appendix, we present the explicit expressions for the perturbative constraints and Hamiltonian that complete the discussion in Sec. \ref{sec: II}. We use the same notation employed in the main text of this article. For the axial sector, the perturbative constraint takes the form
\begin{equation}
    \label{eq: appA-A.1}
    C_0[k_0^{\mathfrak{n},\lambda}] = \sum_{\mathfrak{n},\lambda} k_0^{\mathfrak{n},-\lambda}\bigg[ \frac{(l+2)!}{(l-2)!}(\Omega_b + \Omega_c)\frac{h_2^{\mathfrak{n},-\lambda}}{p_c^2} - 2p_2^{\mathfrak{n},-\lambda} + \lambda\omega_n\bigg( p_1^{\mathfrak{n},\lambda} - 4l(l+1)\frac{L_o^2}{p_b^2}\Omega_bh_1^{\mathfrak{n},\lambda} \bigg)\bigg],
\end{equation}
where $k_0^{\mathfrak{n},\lambda}$ is the corresponding perturbative Lagrange multiplier. The minus sign in front of the $\lambda$--indices indicates the opposite sign (i.e., if $\lambda = +$, then $-\lambda = -$) and arises from derivatives with respect to $\zeta$. As the perturbative variables are refined, these contributions vanish. Conversely, the axial Hamiltonian is given by
\begin{equation}
    \label{eq: appA-A.2}
    \begin{aligned}
        \kappa\tilde{H}^{\text{ax}}[\tilde{N}] &= \sum_{\mathfrak{n},\lambda}\frac{\tilde{N}}{2} \bigg[\frac{p_b^2}{L_o^2}\frac{[p_1^{\mathfrak{n},\lambda}]^2}{l(l+1)} + \bigg(6\Omega_b^2 + 4\Omega_b\Omega_c + \frac{p_b^2}{L_o^2}l(l+1)\bigg)\frac{L_o^2}{p_b^2}l(l+1)[h_1^{\mathfrak{n},\lambda}]^2 - 4\Omega_b h_1^{\mathfrak{n},\lambda}p_1^{\mathfrak{n},\lambda}  - 4\Omega_ch_2^{\mathfrak{n},\lambda}p_2^{\mathfrak{n},\lambda}\\
        &+ 4p_c^2\frac{(l-2)!}{(l+2)!}[p_2^{\mathfrak{n},\lambda}]^2 + \bigg(2\Omega_b^2 + 4\Omega_c^2 + 4\Omega_b\Omega_c + \omega_n^2p_c^2\bigg)\frac{1}{4p_c^2}\frac{(l+2)!}{(l-2)!}[h_2^{\mathfrak{n},\lambda}]^2 + \lambda\omega_n\frac{(l+2)!}{(l-2)!}h_1^{\mathfrak{n},\lambda}h_2^{\mathfrak{n},-\lambda}\bigg].
    \end{aligned}
\end{equation}
We note that, in Ref. \cite{MM}, the densitized lapse function was defined with a normalization that differs by a factor of $2$ with respect to the one adopted here. This accounts for the overall relative factor of $1/2$ appearing in our expression, compared to that in Ref. \cite{MM}. 

Similarly, the perturbative constraints for the polar sector can be formulated as
\begin{equation}
    \label{eq: appA-A.3}
    \begin{aligned}
        &\begin{aligned}
            C_1[k_1^{\mathfrak{n},\lambda}] &= \sum_{\mathfrak{n},\lambda} k_1^{\mathfrak{n},\lambda}\bigg[l(l+1)\bigg(\lambda\omega_n|p_c|\bigg[4\frac{L_o^2}{p_b^2}\Omega_bh_5^{\mathfrak{n},-\lambda} - \frac{p_5^{\mathfrak{n},-\lambda}}{l(l+1)}\bigg]- 2\frac{L_o^2}{p_b^2}|p_c|\Omega_b h_6^{\mathfrak{n},\lambda} - |p_c|p_3^{\mathfrak{n},\lambda}\bigg) + 2|p_c|p_4^{\mathfrak{n},\lambda}\\
            &- \frac{(l+2)!}{(l-2)!}(\Omega_b+\Omega_c)\frac{h_4^{\mathfrak{n},\lambda}}{|p_c|}\bigg],
        \end{aligned}\\
        &\begin{aligned}
            C_2[k_2^{\mathfrak{n},\lambda}] &= \sum_{\mathfrak{n},\lambda} k_2^{\mathfrak{n},\lambda}\bigg[2\lambda\omega_n|p_c|\bigg(\frac{L_o^2}{p_b^2}\Omega_b|p_c|h_6^{\mathfrak{n},-\lambda} - (\Omega_b+\Omega_c)\frac{h_3^{\mathfrak{n},-\lambda}}{|p_c|} - \frac{p_b^2}{L_o^2}\frac{p_6^{\mathfrak{n},-\lambda}}{|p_c|}\bigg) - 2l(l+1)(\Omega_b+\Omega_c)h_5^{\mathfrak{n},\lambda}\\
            & + \frac{p_b^2}{L_o^2}p_5^{\mathfrak{n},\lambda}\bigg],
        \end{aligned}\\
&\begin{aligned}
            C_3[k_3^{\mathfrak{n},\lambda}] &= -\sum_{\mathfrak{n},\lambda} \tilde{N}k_3^{\mathfrak{n},\lambda}\bigg[\Omega_b|p_c|p_3^{\mathfrak{n},\lambda} - \bigg(\frac{l^2+l+2}{2}\frac{p_b^2}{L_o^2} + \omega_n^2p_c^2 + 2\Omega_b^2 + 2\Omega_b\Omega_c + 4\frac{\tilde{H}_{\text{KS}}}{L_o}\bigg)\frac{h_3^{\mathfrak{n},\lambda}}{|p_c|} + \frac{p_b^2}{L_o^2}\Omega_c\frac{p_6^{\mathfrak{n},\lambda}}{|p_c|}\\
            &- \frac{L_o^2}{p_b^2}\bigg(2\Omega_b\Omega_c + \frac{l^2+l+2}{2}\frac{p_b^2}{L_o^2} + 2\frac{\tilde{H}_{\text{KS}}}{L_o}\bigg)|p_c|h_6^{\mathfrak{n},\lambda} + \lambda\omega_n|p_c|l(l+1)h_5^{\mathfrak{n},-\lambda} - \frac{1}{4}\frac{(l+2)!}{(l-2)!}\frac{p_b^2}{L_o^2}\frac{h_4^{\mathfrak{n},\lambda}}{|p_c|}\bigg],
        \end{aligned}
\end{aligned}
\end{equation}
where $k_1^{\mathfrak{n},\lambda}$, $k_2^{\mathfrak{n},\lambda}$, and $k_3^{\mathfrak{n},\lambda}$ denote their corresponding perturbative Lagrange multipliers. On the other hand, the perturbative polar Hamiltonian is given by
\begin{equation}
    \label{eq: appA-A.4}
    \begin{aligned}
        \kappa \tilde{H}^{\text{po}}[\tilde{N}] &= \sum_{\mathfrak{n},\lambda}\frac{\tilde{N}}{2}\bigg[  \bigg(8\Omega_b(\Omega_b+\Omega_c)+4\frac{p_b^2}{L_o^2}-\omega_n^2p_c^2+8\frac{\tilde{H}_{\text{KS}}}{L_o}\bigg)\frac{[h_3^{\mathfrak{n},\lambda}]^2}{p_c^2} - (4\Omega_b-2\Omega_c)h_6^{\mathfrak{n},\lambda}p_6^{\mathfrak{n},\lambda} + 2\frac{L_o^2}{p_b^2} p_c^2 \Omega_b h_6^{\mathfrak{n},\lambda}p_3^{\mathfrak{n},\lambda}\\
        &+ \frac{p_b^4}{L_o^4}\frac{[p_6^{\mathfrak{n},\lambda}]^2}{p_c^2} + \bigg(4\Omega_b(\Omega_b+\Omega_c)+2\frac{p_b^2}{L_o^2}+3\frac{\tilde{H}_{\text{KS}}}{L_o}\bigg)\frac{L_o^4}{p_b^4}p_c^2[h_6^{\mathfrak{n},\lambda}]^2 + 2\frac{p_b^2}{L_o^2}\bigg(\frac{2}{p_c^2}\Omega_bh_3^{\mathfrak{n},\lambda} - p_3^{\mathfrak{n},\lambda}\bigg)p_6^{\mathfrak{n},\lambda}\\
        &- \frac{L_o^2}{p_b^2}\bigg(4\Omega_b(\Omega_b-\Omega_c)+(l+2)(l-1)\frac{p_b^2}{L_o^2}-4\frac{\tilde{H}_{\text{KS}}}{L_o}\bigg)h_3^{\mathfrak{n},\lambda}h_6^{\mathfrak{n},\lambda} + 2\lambda\omega_nl(l+1)h_3^{\mathfrak{n},\lambda}h_5^{\mathfrak{n},-\lambda} + \frac{p_b^2}{L_o^2}\frac{[p_5^{\mathfrak{n},\lambda}]^2}{l(l+1)}\\
        &+ 2\frac{L_o^2}{p_b^2}l(l+1)\bigg(4\Omega_b(\Omega_b+\Omega_c)+\frac{p_b^2}{L_o^2}+2\frac{\tilde{H}_{\text{KS}}}{L_o}\bigg)[h_5^{\mathfrak{n},\lambda}]^2 - 4\Omega_bh_5^{\mathfrak{n},\lambda}p_5^{\mathfrak{n},\lambda} + 4p_c^2\frac{(l-2)!}{(l+2)!}[p_4^{\mathfrak{n},\lambda}]^2\\
        &+ \frac{1}{4p_c^2}\frac{(l+2)!}{(l-2)!}\bigg(4(\Omega_b+\Omega_c)^2+2\frac{p_b^2}{L_o^2}+\omega_n^2p_c^2+4\frac{\tilde{H}_{\text{KS}}}{L_o}\bigg)[h_4^{\mathfrak{n},\lambda}]^2 - \frac{1}{2}\frac{(l+2)!}{(l-2)!}h_6^{\mathfrak{n},\lambda}h_4^{\mathfrak{n},\lambda} - 4\Omega_ch_4^{\mathfrak{n},\lambda}p_4^{\mathfrak{n},\lambda}\\
        &- \frac{(l+2)!}{(l-2)!}\lambda\omega_nh_5^{\mathfrak{n},\lambda}h_4^{\mathfrak{n},-\lambda}\bigg].
    \end{aligned}
\end{equation}

Throughout the entire analysis carried out in this work (not only in Sec. \ref{sec: II}), the background variables and the background Lagrange multipliers must receive suitable quadratic perturbative corrections in order to preserve the canonical Hamiltonian structure of the total constrained system describing background and perturbations. In the case of the background phase-space variables, this is discussed in Appendix \ref{app: D}. Regarding the formulas for Sec. \ref{sec: II} presented in this appendix, the lapse function is the only quantity affected. Its correction is required to render the set of perturbative constraints Abelian, or equivalently, to ensure that their Poisson brackets vanish identically \cite{LMM,MM2}. Since further corrections to the lapse are introduced in Sec. \ref{sec: III}, and intermediate expressions are not especially illuminating, we display here only its contribution to the total correction. The full expression will be presented once all corrections have been specified. Thus, we must replace $\tilde{N}$ with $(1 + \tilde{n}_c)\tilde{N}$, where the relative correction $\tilde{n}_c$ is a second-order perturbative quantity given by
\begin{equation}
    \label{eq: appA-A.5}
    \tilde{n}_c = - \sum_{\mathfrak{n},\lambda}\frac{k_3^{\mathfrak{n},\lambda}}{L_o|p_c|}\bigg[\frac{L_o^2}{p_b^2}p_c^2h_6^{\mathfrak{n},\lambda}+ 2h_3^{\mathfrak{n},\lambda}\bigg].
\end{equation}

\section{Computation of the perturbative gauge invariants \label{app: B}}

In this appendix, we present the intermediate steps omitted in the Hamiltonian discussion of Sec. \ref{sec: III}, which lead to the construction of the perturbative gauge invariants. Again, we use the same notation as in the main text.

For the axial sector, we consider the canonical transformation,
\begin{equation}
    \label{eq: appB-B.1}
    \begin{aligned}
        &h_1^{\mathfrak{n,\lambda}} = P_1^{\mathfrak{n,\lambda}} - \frac{\lambda\omega_n}{2}Q_2^{\mathfrak{n,-\lambda}},\\
        &h_2^{\mathfrak{n,\lambda}} = Q_2^{\mathfrak{n,\lambda}},\\
        &p_1^{\mathfrak{n,\lambda}} = -Q_1^{\mathfrak{n,\lambda}} - 2\lambda\omega_nl(l+1)\frac{L_o^2}{p_b^2}\Omega_bQ_2^{\mathfrak{n,-\lambda}},\\
        &p_2^{\mathfrak{n,\lambda}} = P_2^{\mathfrak{n,\lambda}} + \frac{\lambda\omega_n}{2}Q_1^{\mathfrak{n,-\lambda}} + 2\lambda\omega_nl(l+1)\frac{L_o^2}{p_b^2}\Omega_bP_1^{\mathfrak{n,-\lambda}} + \frac{1}{2}\frac{(l+2)!}{(l-2)!}\frac{1}{p_c^2}(\Omega_b+\Omega_c)Q_2^{\mathfrak{n,\lambda}},
    \end{aligned}
\end{equation}
under which $P_2^{\mathfrak{n},\lambda}$ becomes the perturbative axial constraint. Because the transformation is background dependent, the perturbative Hamiltonian acquires a new form that cannot be obtained by simply replacing in its original form the relation between the original and the transformed variables \cite{LMM,MM}. The final expression is analogous to the axial counterpart of Eq. \eqref{eq: III-3.3}, but with additional contributions. These extra terms are either proportional to the axial perturbative constraint or to the background Hamiltonian. In practice, they are irrelevant, since they can be eliminated by appropriately redefining the associated (perturbative or background) Lagrange multipliers. Schematically, using the notation $F' = \{F,\tilde{H}_{\text{KS}}\}_{\text{B}}$, the computation after the canonical transformation yields:
\begin{equation}
    \label{eq: appB-B.2}
    \mathbf{\tilde{H}}^{\text{ax}} = \tilde{H}^{\text{ax}} + \sum_{\mathfrak{n},\lambda}\sum_{i=1}^2\big[ P_i^{\mathfrak{n},\lambda}\big(Q_i^{\mathfrak{n},\lambda}\big)' - p_i^{\mathfrak{n},\lambda}\big(h_i^{\mathfrak{n},\lambda}\big)'\big] - \mathbf{h}^{\text{ax}}\big[\tilde{H}_{\text{KS}},P_2^{\mathfrak{n},\lambda}\big],
\end{equation}
where $\mathbf{h}^{\text{ax}}$ denotes the Hamiltonian term proportional to the constraints that appear as its arguments. The remaining term reflects the fact that the Hamiltonians are not strictly identical under a background-dependent canonical transformation. By redefining the perturbative Lagrange multiplier as
\begin{equation}
    \label{eq: appB-B.3}
    \kappa\textbf{h}_0^{\mathfrak{n},\lambda} = -2\kappa h_0^{\mathfrak{n},\lambda} + 2\tilde{N}p_c^2\frac{(l-2)!}{(l+2)!}\left[P_2^{\mathfrak{n},\lambda} + \lambda\omega_nQ_1^{\mathfrak{n},-\lambda} + 4\lambda\omega_nl(l+1)\frac{L_o^2}{p_b^2}\Omega_bP_1^{\mathfrak{n},-\lambda} + \frac{1}{p_c^2}\frac{(l+2)!}{(l-2)!}\Omega_bQ_2^{\mathfrak{n},\lambda}\right],
\end{equation}
and adjusting the lapse function through the relative modification,
\begin{equation}
    \label{eq: appB-B.4}
    \tilde{n}^{\text{ax}} = \frac{1}{2\kappa L_o}\sum_{\mathfrak{n},\lambda}l(l+1)\frac{p_b^2}{L_o^2}[P_1^{\mathfrak{n},\lambda}]^2,
\end{equation}
the $\mathbf{h}^{\text{ax}}$ contribution is removed from the Hamiltonian. The resulting Hamiltonian coefficients are
\begin{equation}
    \label{eq: appB-B.5}
    \begin{aligned}
        &\begin{aligned}
            A_{\text{ax}} = \frac{1}{l(l+1)}\frac{p_b^2}{L_o^2}\bigg[1+\frac{\omega_n^2p_c^2}{(l+2)(l-1)}\frac{L_o^2}{p_b^2}\bigg], 
        \end{aligned}\\
        &\begin{aligned}
            B_{\text{ax}} = l(l+1)\frac{L_o^2}{p_b^2}\bigg[8\Omega_b^2 + 8\Omega_b\Omega_c + (l^2+l+2)\frac{p_b^2}{L_o^2} + \frac{16\omega_n^2p_c^2}{(l+2)(l-1)}\frac{L_o^2}{p_b^2}\Omega_b^2\bigg],
        \end{aligned}\\
        &\begin{aligned}
            C_{\text{ax}} = 4\Omega_b\bigg[1+2\frac{\omega_n^2p_c^2}{(l+2)(l-1)}\frac{L_o^2}{p_b^2}\bigg],
        \end{aligned}
    \end{aligned}
\end{equation}
completing the definition of $\mathbf{\tilde{H}}^{\text{ax}}$. To ensure a consistent density weight throughout the full Hamiltonian expression, the axial contribution has been slightly rearranged with respect to the form presented in Ref. \cite{MM}.

Similarly, for the polar sector we consider the following canonical transformation of the configuration variables:
\begin{equation}
    \label{eq: appB-B.6}
    \begin{aligned}
        &h_3^{\mathfrak{n,\lambda}} = |p_c|\bigg[Q_3^{\mathfrak{n},\lambda} + \Omega_bQ_4^{\mathfrak{n},\lambda} -\frac{l(l+1)}{2}Q_5^{\mathfrak{n},\lambda}\bigg],\\
        &h_4^{\mathfrak{n,\lambda}} = |p_c|Q_5^{\mathfrak{n},\lambda},\\
        &h_5^{\mathfrak{n,\lambda}} = \frac{p_b^2}{L_o^2}Q_6^{\mathfrak{n},\lambda} + \lambda\omega_n\frac{|p_c|}{2}Q_5^{\mathfrak{n},-\lambda},\\
        &h_6^{\mathfrak{n,\lambda}} = \frac{p_b^2}{L_o^2}\frac{1}{|p_c|}\bigg[\Omega_cQ_4^{\mathfrak{n},\lambda} + \frac{2}{\Lambda}\frac{L_o^2}{p_b^2}(\Omega_bP_3^{\mathfrak{n},\lambda} - P_4^{\mathfrak{n},\lambda}) + 2\lambda\omega_n|p_c|Q_6^{\mathfrak{n},-\lambda}\bigg],
    \end{aligned}
\end{equation}
where $\Lambda = (l+2)(l-1) - 6\Omega_b\Omega_cL_o^2/p_b^2$ plays a key role in the identification with the BH master functions. The corresponding relations for the conjugate momenta complete the transformation,
\begin{equation}
    \label{eq: appB-B.7}
    \begin{aligned}
        &\begin{aligned}
            p_3^{\mathfrak{n,\lambda}} &= \frac{1}{|p_c|}\bigg[\bigg(1-\frac{4}{\Lambda}\frac{L_o^2}{p_b^2}\Omega_b^2\bigg)P_3^{\mathfrak{n,\lambda}} + \frac{1}{\Omega_b}\bigg(\frac{1}{2}(l+2)(l-1)\frac{p_b^2}{L_o^2} + \omega_n^2p_c^2\bigg)Q_3^{\mathfrak{n,\lambda}} + \bigg(\frac{\Lambda}{2}\frac{p_b^2}{L_o^2} + \omega_n^2p_c^2 + \Omega_b\Omega_c\bigg)Q_4^{\mathfrak{n,\lambda}}\\
            &+\frac{4}{\Lambda}\frac{L_o^2}{p_b^2}\Omega_b P_4^{\mathfrak{n,\lambda}} - 2\lambda\omega_n|p_c|(\Omega_b+\Omega_c)Q_6^{\mathfrak{n,-\lambda}}\bigg],
        \end{aligned}\\
        &\begin{aligned}
            p_4^{\mathfrak{n,\lambda}} &= \frac{1}{|p_c|}P_5^{\mathfrak{n,\lambda}} - 2\lambda\omega_n(\Omega_b+\Omega_c)Q_6^{\mathfrak{n,-\lambda}} + \bigg[\frac{l(l+1)}{2|p_c|}\frac{1}{\Omega_b}\bigg(\frac{1}{2}(l+2)(l-1)\frac{p_b^2}{L_o^2} + \omega_n^2p_c^2\bigg) - (\Omega_b-\Omega_c)\bigg]Q_3^{\mathfrak{n,\lambda}}\\
            &+ \frac{\lambda\omega_n}{2}\frac{L_o^2}{p_b^2}P_6^{\mathfrak{n,-\lambda}} + \frac{L_o^2}{p_b^2}\frac{1}{|p_c|}\bigg[\omega_n^2p_c^2\bigg(\Omega_b^2-\Omega_b\Omega_c-\frac{\Lambda}{2}\frac{p_b^2}{L_o^2}\bigg)+\frac{l(l+1)}{2}\frac{p_b^2}{L_o^2}\bigg(\frac{1}{2}(l+2)(l-1)\frac{p_b^2}{L_o^2} + \omega_n^2p_c^2\bigg)\bigg]Q_4^{\mathfrak{n,\lambda}}\\
            &-\bigg[1+\omega_n^2|p_c|\frac{L_o^2}{p_b^2}\bigg]\frac{2}{\Lambda}\frac{L_o^2}{p_b^2}\Omega_b P_4^{\mathfrak{n},\lambda} + \bigg[\frac{l(l+1)}{2|p_c|} + \bigg(1+\omega_n^2|p_c|\frac{L_o^2}{p_b^2}\bigg)\frac{2}{\Lambda}\frac{L_o^2}{p_b^2}\Omega_b^2\bigg]P_3^{\mathfrak{n},\lambda} + \frac{(l+2)!}{(l-2)!}\frac{\Omega_b+\Omega_c}{2|p_c|}Q_5^{\mathfrak{n},\lambda},
        \end{aligned}\\
        &\begin{aligned}
            p_5^{\mathfrak{n,\lambda}} &= \frac{L_o^2}{p_b^2}P_6^{\mathfrak{n,\lambda}} - 2\lambda\omega_n|p_c|\frac{L_o^2}{p_b^2}\bigg[(\Omega_b-\Omega_c)Q_3^{\mathfrak{n,-\lambda}} - \bigg(\frac{2}{\Lambda}\frac{L_o^2}{p_b^2} - \Omega_b^2 + \Omega_b\Omega_c\bigg)Q_4^{\mathfrak{n,-\lambda}} - l(l+1)\Omega_bQ_5^{\mathfrak{n,-\lambda}}\bigg]\\
            &+ 2(l+2)(l-1)(\Omega_b+\Omega_c)Q_6^{\mathfrak{n,\lambda}},
        \end{aligned}\\
        &\begin{aligned}
            p_6^{\mathfrak{n,\lambda}} &= \frac{L_o^2}{p_b^2}|p_c|\bigg[\bigg(\frac{\Lambda}{2}\frac{p_b^2}{L_o^2} + \Omega_b\Omega_c - 2\Omega_b^2\bigg)Q_4^{\mathfrak{n},\lambda} - 2\Omega_bQ_3^{\mathfrak{n},\lambda} + \frac{2}{\Lambda}\frac{L_o^2}{p_b^2}\Omega_b(\Omega_bP_3^{\mathfrak{n},\lambda} - P_4^{\mathfrak{n},\lambda}) + 2\lambda\omega_n|p_c|\Omega_bQ_6^{\mathfrak{n},-\lambda}\\
            &+ l(l+1)\Omega_bQ_5^{\mathfrak{n},\lambda}\bigg].
        \end{aligned}
    \end{aligned}
\end{equation}

After this transformation, $P_4^{\mathfrak{n},\lambda}$, $P_5^{\mathfrak{n},\lambda}$, and $P_6^{\mathfrak{n},\lambda}$ become the perturbative polar constraints. In the polar sector, however, an additional redefinition of the third constraint is required, analogous to the one introduced in the previous appendix. It is given by
\begin{equation}
    \label{eq: appB-B.8}
    \mathbf{C}_3[k_3^{\mathfrak{n},\lambda}] = C_3[k_3^{\mathfrak{n},\lambda}] + \sum_{\mathfrak{n},\lambda}2\tilde{N}k_3^{\mathfrak{n},\lambda}\bigg[2Q_3^{\mathfrak{n},\lambda} + (2\Omega_b+\Omega_c)Q_4^{\mathfrak{n},\lambda} + \frac{2}{\Lambda}\frac{L_o^2}{p_b^2}(\Omega_bP_3^{\mathfrak{n},\lambda} - P_4^{\mathfrak{n},\lambda})\bigg]\frac{\tilde{H}_{\text{KS}}}{L_o},
\end{equation}
and induces a correction in the densitized lapse function corresponding to the relative contribution,
\begin{equation}
    \label{eq: appB-B.9}
    \tilde{n}^c = -\sum_{\mathfrak{n},\lambda}\frac{2k_3^{\mathfrak{n},\lambda}}{L_o}\bigg[2Q_3^{\mathfrak{n},\lambda} + (2\Omega_b+\Omega_c)Q_4^{\mathfrak{n},\lambda} + \frac{2}{\Lambda}\frac{L_o^2}{p_b^2}(\Omega_bP_3^{\mathfrak{n},\lambda} - P_4^{\mathfrak{n},\lambda})\bigg].
\end{equation}
As in the axial case, the perturbative polar Hamiltonian acquires a new structure owing to the background dependence of the transformation. After performing the canonical transformation, the Hamiltonian can be decomposed according to the same criterion, leading to
\begin{equation}
    \label{eq: appB-B.10}
    \mathbf{\tilde{H}}^{\text{po}} = \tilde{H}^{\text{po}} + \sum_{\mathfrak{n},\lambda}\sum_{i=3}^6\big[ P_i^{\mathfrak{n},\lambda}\big(Q_i^{\mathfrak{n},\lambda}\big)' - p_i^{\mathfrak{n},\lambda}\big(h_i^{\mathfrak{n},\lambda}\big)'\big] - \mathbf{h}^{\text{po}}\big[\tilde{H}_{\text{KS}},P_4^{\mathfrak{n},\lambda},P_5^{\mathfrak{n},\lambda},P_6^{\mathfrak{n},\lambda}\big],
\end{equation}
where $\mathbf{h}^{\text{po}}$ stands for the Hamiltonian terms proportional to constraints. These contributions can be removed by redefining the perturbative Lagrange multipliers as follows:
\begin{equation}
    \label{eq: appB-B.11}
    \begin{aligned}
        &\begin{aligned}
            \kappa\mathbf{k}_1^{\mathfrak{n},\lambda} &= 2\kappa k_1^{\mathfrak{n},\lambda} + 4\tilde{N}\frac{(l-2)!}{(l+2)!}\frac{L_o^2}{p_b^2}\bigg[\omega_n^2p_c^2\bigg((\Omega_b-\Omega_c)Q_3^{\mathfrak{n},\lambda}-\bigg(\frac{\Lambda}{2}\frac{p_b^2}{L_o^2} + \Omega_b\Omega_c - \Omega_b^2\bigg)Q_4^{\mathfrak{n},\lambda}\bigg)\\
            &+ \frac{l(l+1)}{4}\frac{p_b^2}{L_o^2}\bigg(\bigg(1+\frac{4}{\Lambda}\frac{p_b^2}{L_o^2}\Omega_b^2\bigg)P_3^{\mathfrak{n},\lambda} + \bigg(\frac{1}{2}(l+2)(l-1)\frac{p_b^2}{L_o^2} + \omega_n^2p_c^2\bigg)\frac{1}{\Omega_b}Q_3^{\mathfrak{n},\lambda}\\
            &+ \bigg(\frac{\Lambda}{2}\frac{p_b^2}{L_o^2} + \omega_n^2p_c^2 + 5\Omega_b\Omega_c\bigg)Q_4^{\mathfrak{n},\lambda} - \frac{4}{\Lambda}\frac{p_b^2}{L_o^2}\Omega_b P_4^{\mathfrak{n},\lambda} + \frac{2P_5^{\mathfrak{n},\lambda}}{l(l+1)}\bigg) + \frac{1}{2}\lambda\omega_n|p_c|P_6^{\mathfrak{n},-\lambda}\bigg],
        \end{aligned}\\
        &\begin{aligned}
            \kappa\mathbf{k}_2^{\mathfrak{n},\lambda} &= \kappa k_2^{\mathfrak{n},\lambda} - \frac{\tilde{N}}{2}\bigg[4\lambda\omega_n|p_c|\frac{(l-2)!}{(l+2)!}\frac{L_o^4}{p_b^4}\bigg(\bigg[(l+2)(l-1)\frac{p_b^2}{L_o^2}+\omega_n^2p_c^2\bigg]\bigg[(\Omega_b-\Omega_c)Q_3^{\mathfrak{n},-\lambda}\\
            &- \bigg(\frac{\Lambda}{2}\frac{p_b^2}{L_o^2} + \Omega_b\Omega_c - \Omega_b^2\bigg)Q_4^{\mathfrak{n},-\lambda}\bigg] + \frac{1}{2}l(l+1)\frac{p_b^2}{L_o^2}\bigg[\bigg(\frac{1}{2}(l+2)(l-1)\frac{p_b^2}{L_o^2}+\omega_n^2p_c^2\bigg)\\
            &\times \left(Q_4^{\mathfrak{n},-\lambda} +\frac{1}{\Omega_b} Q_3^{\mathfrak{n},-\lambda}\right) + P_3^{\mathfrak{n},-\lambda}\bigg] + \frac{1}{2}(l+2)(l-1)\frac{p_b^2}{L_o^2}\bigg[\bigg(\frac{\Lambda}{2}\frac{p_b^2}{L_o^2} + \Omega_b\Omega_c - 2\Omega_b^2\bigg)Q_4^{\mathfrak{n},-\lambda}\\
            &- 2\Omega_bQ_3^{\mathfrak{n},-\lambda} + \frac{2}{\Lambda}\frac{L_o^2}{p_b^2}\Omega_b(\Omega_bP_3^{\mathfrak{n},-\lambda} - P_4^{\mathfrak{n},-\lambda} )\bigg]\bigg) + 4\Omega_bQ_6^{\mathfrak{n},\lambda}\\
            &- \frac{(l-2)!}{(l+2)!}\frac{L_o^4}{p_b^4}\bigg((l+2)(l-1)\frac{p_b^2}{L_o^2}+\omega_n^2p_c^2\bigg)P_6^{\mathfrak{n},\lambda}\bigg],
        \end{aligned}\\
        &\begin{aligned}
            \kappa\mathbf{k}_3^{\mathfrak{n},\lambda} &= -\kappa \tilde{N}k_3^{\mathfrak{n},\lambda} - \tilde{N}(\Omega_b+\Omega_c)Q_4^{\mathfrak{n},\lambda} + \tilde{N}\frac{(l+2)(l-1)}{\Lambda}Q_3^{\mathfrak{n},\lambda}\\
            &+ \frac{2}{\Lambda}\frac{L_o^2}{p_b^2}\bigg(2Q_3^{\mathfrak{n},\lambda} + (2\Omega_b+5\Omega_c)Q_4^{\mathfrak{n},\lambda} + \frac{1}{\Lambda}\frac{L_o^2}{p_b^2}(2\Omega_bP_3^{\mathfrak{n},\lambda} + P_4^{\mathfrak{n},\lambda})\bigg)\frac{\tilde{H}_{\text{KS}}[\tilde{N}]}{L_o},
        \end{aligned}
    \end{aligned}
\end{equation}
and by adjusting the lapse with the relative contribution,
\begin{equation}
    \label{eq: appB-B.12}
    \begin{aligned}
        \tilde{n}^{\text{po}} &= \frac{1}{\kappa L_o}\sum_{\mathfrak{n},\lambda}\bigg[2\lambda\omega_n|p_c|\bigg((\Omega_c- 2\Omega_b)Q_4^{\mathfrak{n},\lambda} + \frac{2}{\Lambda}\frac{L_o^2}{p_b^2}(\Omega_bP_3^{\mathfrak{n},\lambda}-P_4^{\mathfrak{n},\lambda}) - 2Q_3^{\mathfrak{n},\lambda}\bigg)Q_6^{\mathfrak{n},-\lambda} + \frac{l(l+1)}{2}\frac{L_o^2}{p_b^2}\\
        &\times\bigg((l+2)(l-1)\frac{p_b^2}{L_o^2}+\omega_n^2p_c^2\bigg)[Q_5^{\mathfrak{n},\lambda}]^2 + l(l+1)Q_5^{\mathfrak{n},\lambda}\bigg(\Omega_cQ_4^{\mathfrak{n},\lambda} + \frac{2}{\Lambda}\frac{L_o^2}{p_b^2}(\Omega_bP_3^{\mathfrak{n},\lambda}-P_4^{\mathfrak{n},\lambda})\bigg)\\
        &+ 2\bigg(l(l+1)\frac{p_b^2}{L_o^2}+\omega_n^2p_c^2\bigg)[Q_6^{\mathfrak{n},\lambda}]^2 - \frac{4}{\Lambda}\frac{L_o^2}{p_b^2}\Omega_bQ_3^{\mathfrak{n},\lambda}P_3^{\mathfrak{n},\lambda} + 2\bigg(\frac{1}{\Lambda}\frac{L_o^2}{p_b^2}\Omega_b\bigg)^2[P_3^{\mathfrak{n},\lambda}]^2 -2\Omega_cQ_3^{\mathfrak{n},\lambda}Q_4^{\mathfrak{n},\lambda}\\
        &-\frac{(l+2)!}{(l-2)!}\frac{L_o^4}{p_b^4}\bigg((l+2)(l-1)\frac{p_b^2}{L_o^2}+\omega_n^2p_c^2\bigg)\bigg(\frac{4}{\Omega_b}\bigg(\frac{1}{2}(l+2)(l-1)\frac{p_b^2}{L_o^2}+\omega_n^2p_c^2\bigg)l(l+1)\frac{p_b^2}{L_o^2}\\
        &+ 8\omega_n^2p_c^2(\Omega_b-\Omega_c)\bigg)Q_3^{\mathfrak{n},\lambda}Q_4^{\mathfrak{n},\lambda} - \bigg((l+2)(l-1)\frac{p_b^2}{L_o^2}+\omega_n^2p_c^2 + \frac{5}{2}\Omega_c^2 + 2\Omega_b\Omega_c\bigg)[Q_4^{\mathfrak{n},\lambda}]^2\\
        &-2\bigg(1+\frac{1}{\Lambda}\frac{L_o^2}{p_b^2}\Omega_b(2\Omega_b+5\Omega_c) + 2\frac{(l+2)!}{(l-2)!}\frac{L_o^4}{p_b^4}\bigg(\frac{1}{2}(l+2)(l-1)\frac{p_b^2}{L_o^2}+\omega_n^2p_c^2\bigg)l(l+1)\frac{p_b^2}{L_o^2}\bigg)Q_4^{\mathfrak{n},\lambda}P_3^{\mathfrak{n},\lambda}\\
        &+8\omega_n^2p_c^2\frac{(l+2)!}{(l-2)!}\frac{L_o^4}{p_b^4}\bigg((l+2)(l-1)\frac{p_b^2}{L_o^2}+\omega_n^2p_c^2\bigg)\frac{\tilde{H}_{\text{KS}}}{L_o}[Q_4^{\mathfrak{n},\lambda}]^2\bigg].
    \end{aligned}
\end{equation}
At this stage, all correction terms for the lapse function, as discussed in Secs. \ref{sec: II} and \ref{sec: III}, have been collected. We can now combine them into a single expression,
\begin{equation}
    \label{eq: appB-B.13}
    \mathbf{\tilde{N}} = ( 1 + \tilde{n}_c + \tilde{n}^c + \tilde{n}^{\text{ax}} + \tilde{n}^{\text{po}}) \tilde{N}.
\end{equation}

We emphasize that corrections to the perturbative Lagrange multipliers must be included explicitly to maintain consistency at the truncation order, although they are not physically relevant. Conversely, corrections to the background variables and to the background densitized lapse are irrelevant for the perturbations, inasmuch as their effect on the dynamics of the perturbations is higher-order and therefore exceeds the order of our perturbative truncation. Thus, for the purposes of the perturbative analysis, using the corrected or uncorrected background expressions is immaterial \cite{MM,LMM}. This also explains in part why we have adopted the same notation for both sets of background variables in our discussion (the set that we have used at every specific stage can be easily deduced from the context).

Finally, since the density weight of the Hamiltonian is already consistent in the polar sector, the coefficients of $\mathbf{\tilde{H}}^{\text{po}}$ are
\begin{equation}
    \label{eq: appB-B.14}
    \begin{aligned}
        &\begin{aligned}
            A_{\text{po}} &= 4\omega_n^2p_c^2\frac{(l-2)!}{(l+2)!}\frac{L_o^4}{p_b^4}(\Omega_b-\Omega_c)^2\bigg((l+2)(l-1)\frac{p_b^2}{L_o^2} + \omega_n^2p_c^2\bigg) + \frac{l(l+1)}{(l+2)(l-1)}\bigg(\frac{1}{2}(l+2)(l-1)\frac{p_b^2}{L_o^2}+ \omega_n^2p_c^2\bigg)^2\\
            &\times \frac{1}{\Omega_b^2}+ 2\omega_n^2p_c^2\bigg(\frac{ p_b^2}{L_o^2} + \frac{2\omega_n^2p_c^2}{(l+2)(l-1)} \bigg)\frac{L_o^2}{p_b^2}\left(1-\frac{\Omega_c}{\Omega_b}\right)+ \bigg[\bigg(\frac{1}{2}(l+2)(l-1)\frac{p_b^2}{L_o^2} + \omega_n^2p_c^2\bigg)\frac{1}{\Omega_b}\bigg]' - \omega_n^2p_c^2 ,
        \end{aligned}\\
        &\begin{aligned}
            B_{\text{po}} &= \frac{l(l+1)}{(l+2)(l-1)},
        \end{aligned}\\
        &\begin{aligned}
            C_{\text{po}} &= l(l+1)\bigg(\frac{p_b^2}{L_o^2} + \frac{2\omega_n^2p_c^2}{(l+2)(l-1)}\bigg) \frac{1}{\Omega_b}+ \frac{4\omega_n^2p_c^2}{(l+2)(l-1)}\frac{L_o^2}{p_b^2}(\Omega_b-\Omega_c) - \frac{2(l+2)(l-1)}{\Lambda}\Omega_b.
        \end{aligned}
    \end{aligned}
\end{equation}
In addition to gathering the canonical transformations to streamline Sec. \ref{sec: III}, the main result presented in this appendix is the reformulation of the polar transformation, allowing for dilation functions in the denominators, which is the primary reason why we have obtained simple Hamiltonian coefficients and achieved in this way a clearer connection with the master functions in comparison with previous works \cite{MM2}.

\section{Adjustment of the potentials \label{app: C}}

In this appendix, we present minor redefinitions omitted in Sec. \ref{sec: IV} when simplifying the Hamiltonian of Eq. \eqref{eq: III-3.3}. As with previous canonical transformations, Eqs. \eqref{eq: III-3.4} and \eqref{eq: III-3.5} introduce modifications to the Hamiltonian owing to the background dependence. These ultimately yield a diagonal Hamiltonian with unit coefficients multiplying the squared momentum terms, along with a redundant contribution proportional to the background Hamiltonian. Following the reasoning in Appendix \ref{app: B}, this last term can be removed by introducing a quadratic (relative) correction in the lapse function of the form
\begin{equation}
    \label{eq: appC-C.1}
    \tilde{n}^{\text{GS}} =  \frac{1}{\kappa L_o}\sum_{\mathfrak{n},\lambda}\bigg[\frac{4}{\Lambda}\frac{L_o^2}{p_b^2}\bigg(\frac{3}{\Lambda}\bigg[2(l+2)(l-1) - 3\Omega_b\Omega_c\frac{L_o^2}{p_b^2}\bigg]\Omega_b\Omega_c - \omega_n^2p_c^2\bigg)[\mathcal{Q}_3^{\mathfrak{n},\lambda}]^2 + 3[\mathcal{Q}_1^{\mathfrak{n},\lambda}]^2\bigg],
\end{equation}
which is irrelevant at the perturbative level and naturally combines with Eq. \eqref{eq: appB-B.13}.

\section{Perturbative corrections to the background variables \label{app: D}}

In this appendix we explain how the full canonical structure of the phase space, containing both background and perturbative variables, is preserved throughout the analysis. This is essential for a consistent dynamical treatment of the background, for accounting for backreaction, and for the hybrid quantization. The discussion is kept general so it applies to any point in the main text where a canonical transformation may affect this structure.

In Sec. \ref{sec: II} we introduced a global symplectic structure. Subsequent canonical transformations may appear to break it because they were only specified in the perturbative sector. In particular, if the transformation of the perturbations depends on the background while the background is left unchanged, the symplectic structure is spoiled: the new perturbative variables no longer commute with the old background ones. Restoring the canonical structure requires introducing corresponding perturbative corrections to the background.

For the analysis, we denote abstractly the perturbative variables by {$\{X_J^{\mathfrak{n},\lambda}\} = \{X_{q_J}^{\mathfrak{n},\lambda}, X_{p_J}^{\mathfrak{n},\lambda}\}$ and the background variables as $\{x^j\} = \{x^j_q, x^j_p\}$, where the indices $J$ and $j$ label background and perturbative degrees of freedom, respectively, and the subscripts $q$ and $p$ denote configuration and momentum variables. Together they form a canonical structure,
\begin{equation}
    \label{eq: appC-D.1}
    \{X_{q_J}^{\mathfrak{n},\lambda}, X_{p_{J'}}^{\mathfrak{n}',\lambda'}\} = \delta_{JJ'}\delta_{\mathfrak{n}\mathfrak{n}'}\delta_{\lambda\lambda'}, \qquad \{x^j_q, x^{j'}_p\} = \delta^{jj'},
\end{equation}
with all other Poisson brackets, including mixed ones, vanishing. 

We now consider a canonical transformation that initially acts only on the perturbative variables (see Eqs. \eqref{eq: appB-B.1}, \eqref{eq: appB-B.6}, \eqref{eq: appB-B.7}, \eqref{eq: III-3.4}, \eqref{eq: III-3.5}, and \eqref{eq: III-3.9}). The transformed perturbative variables are $\{W_J^{\mathfrak{n},\lambda}\}$, given by means of the relations $ X_J^{\mathfrak{n},\lambda}= X_J^{\mathfrak{n},\lambda}[x^j,W_{J'}^{\mathfrak{n}',\lambda'}]$. While Eq. \eqref{eq: appC-D.1} continues to hold within the perturbative and nonperturbative sectors, the mixed Poisson brackets no longer vanish. This issue is resolved by redefining the background variables as $w^j = w^j[x^{j'},X_J^{\mathfrak{n},\lambda}]$ through the perturbative corrections \cite{LMM},
\begin{equation}
    \label{eq: appC-D.2}
     w^j_{q} = x^j_{q} + \frac{1}{2}\sum_{\mathfrak{n},\lambda}\sum_J \left[X^{\mathfrak{n},\lambda}_{q_J}\frac{\partial X^{\mathfrak{n},\lambda}_{p_J}}{\partial x^j_{p}} - \frac{\partial X^{\mathfrak{n},\lambda}_{q_J}}{\partial x^j_{p}}X^{\mathfrak{n},\lambda}_{p_J}\right], \qquad w^j_{p} = x^j_{p} - \frac{1}{2}\sum_{\mathfrak{n},\lambda}\sum_J \left[X^{\mathfrak{n},\lambda}_{q_J}\frac{\partial X^{\mathfrak{n},\lambda}_{p_J}}{\partial x^j_{q}} - \frac{\partial X^{\mathfrak{n},\lambda}_{q_J}}{\partial x^j_{q}}X^{\mathfrak{n},\lambda}_{p_J}\right],
\end{equation}
where the initial perturbative variables are understood as functions of the new ones when evaluating the partial derivatives. In this way, the symplectic structure is preserved at the truncated order,
\begin{equation}
    \label{eq: appC-D.3}
    \{W_{q_J}^{\mathfrak{n},\lambda}, W_{p_{J'}}^{\mathfrak{n}',\lambda'}\} = \delta_{JJ'}\delta_{\mathfrak{n}\mathfrak{n}'}\delta_{\lambda\lambda'}, \qquad \{w^j_q, w^{j'}_p\} = \delta^{jj'},
\end{equation}
with all other Poisson brackets, including mixed ones, vanishing.

\end{document}